\documentclass[11pt,a4paper]{article}

\usepackage{amsmath}
\usepackage{amssymb}
\usepackage{graphicx}
\usepackage{url}
\usepackage{epsfig}
\usepackage[T1]{fontenc}
\usepackage[latin9]{inputenc}
\usepackage{multirow}
\newcommand{\tb}{\bar t}

\newcommand{\tth}{ t \tb H}
\newcommand{\ttW}{ t \tb W}
\newcommand{\ttV}{ t \tb V}
\newcommand{\ttWp}{ t \tb W^{+}}
\newcommand{\ttWm}{ t \tb W^{-}}

\newcommand{\ttZ}{ t \tb Z}
\newcommand{\al}{\alpha}
\newcommand{\als}{\alpha_{\rm s}}
\newcommand{\shat}{\hat s}

\newcommand{\muf}{\mu_{\rm F}}
\newcommand{\mur}{\mu_{\rm R}}
\newcommand{\mufo}{\mu_{{\rm F},0}}
\newcommand{\muro}{\mu_{{\rm R},0}}

\newcommand{\sigh}{\hat \sigma}

\newcommand{\nn}{\nonumber}

\newcommand{\tosv}{{\scriptscriptstyle \to}}

\RequirePackage[numbers,sort&compress]{natbib}

\def\beq{\begin{equation}}
\def\eeq{\end{equation}}
\def\bear{\begin{eqnarray}}
\def\eear{\end{eqnarray}}

\def\bet34{\beta_{kl}}
\def\CF{C_{\mathrm{F}}}
\def\CA{C_{\mathrm{A}}}
\def\nf{n_{\mathrm{f}}}
\def\TR{T_{\mathrm{R}}}

\begin{document}

\begin{flushright}
	MS-TP-18-29
\end{flushright}
\vspace{1cm}

\begin{center}
	{\Large\textbf{Associated production of a top quark pair \\[0.5ex] with a heavy electroweak
gauge boson \\[0.5ex] at NLO+NNLL accuracy\\ \mbox{}}}\\
	\vspace{.7cm}
	Anna Kulesza$^{a,}$\footnote{\texttt{anna.kulesza@uni-muenster.de}}, Leszek Motyka$^{b,}$\footnote{\texttt{leszekm@th.if.uj.edu.pl}}, Daniel Schwartl\"ander$^{a,}$\footnote{\texttt{d\_schw20@uni-muenster.de}}, Tomasz Stebel$^{c,d,}$\footnote{\texttt{tomasz.stebel@uj.edu.pl}} and Vincent Theeuwes$^{e,f,}$\footnote{\texttt{vincent.theeuwes@uni-goettingen.de}}

	\vspace{.3cm}
	\textit{
		$^a$ Institute for Theoretical Physics, WWU M\"unster, D-48149 M\"unster, Germany\\
		$^b$ Institute of Physics, Jagiellonian University, S.\L{}ojasiewicza 11, 30-348 Krak\'ow, Poland\\
		$^c$ Institute of Nuclear Physics PAN, Radzikowskiego 152, 31-342 Krak\'ow, Poland\\
			$^d$ Physics Department, Brookhaven National Laboratory, Upton, NY 11973, USA\\
		$^e$ Institute for Theoretical Physics, Georg-August-Univesity G\"ottingen, Friedrich-Hund-Platz 1, 37077 G\"ottingen, Germany\\
		$^f$ Institut de Physique Th\'eorique, Paris Saclay University, CEA, CNRS, F-91191 Gif-sur-Yvette France
	}
\end{center}   

\vspace*{2cm}
\begin{abstract}
We perform threshold resummation of soft gluon corrections to the total cross sections and the invariant mass distributions for production of a top-antitop quark pair associated with a heavy electroweak boson $V = W^+$, $W^-$ or $Z$ in $pp$ collisions at the Large Hadron Collider. The resummation is carried out at next-to-next-to-leading-logarithmic (NNLL) accuracy using the direct QCD Mellin space technique in the three-particle invariant mass kinematics. It is found that for the $t\bar t Z$  process the soft gluon resummation introduces significant corrections to the next-to-leading order (NLO) results. For the central scale equal to the $t\bar t Z$ invariant mass the corrections reach nearly 30\%.  For this process, the dominant theoretical uncertainty of the cross section due to the scale choice is significantly reduced at the NLO+NNLL level with respect to the NLO results. The effects of resummation are found to be less pronounced in the $t\bar t W^{\pm}$ case. The obtained results are compared to recent measurements performed by CMS and ATLAS collaborations at the LHC.

\end{abstract}

\clearpage
\tableofcontents
\setcounter{footnote}{0}

\section{Introduction}
\label{s:intro}

The measurements of associated production of a massive vector boson with a top-antitop quark pair at the LHC~\cite{Chatrchyan:2013qca,Khachatryan:2014ewa,Aad:2015eua,Khachatryan:2015sha,Aaboud:2016xve,Sirunyan:2017uzs,ATLAS:2018ekx} provide an important test of the Standard Model (SM). Together with the associated Higgs boson production with a top-quark pair, they belong to a class of processes with the heaviest final states which can be precisely studied at the LHC. Such studies command particular attention as a means to indirectly search for signals of physics Beyond the Standard Model (BSM).  Additional, they form dominant background in direct BSM searches, as well as to SM measurements, especially the associated Higgs boson production process. It is therefore necessary to know the theoretical predictions for $pp \to \ttV$, $V= W^+, W^-,Z$ with high accuracy.

Over the years, there has been a great effort to improve the theoretical description of the $pp \to \ttV$ process. Next-to-leading-order (NLO) QCD corrections were calculated~\cite{Lazopoulos:2008de, Lazopoulos:2007bv, Hirschi:2011pa, Kardos:2011na, Maltoni:2014zpa, Maltoni:2015ena, Badger:2010mg, Campbell:2012dh, Rontsch:2014cca, Rontsch:2015una} and matched to partons showers~\cite{Garzelli:2012bn, Alwall:2014hca, Maltoni:2015ena}. The electroweak corrections and the combined electroweak corrections with the QCD corrections are also known~\cite{Frixione:2014qaa, Frixione:2015zaa, Frederix:2017wme}. In the light of the full next-to-next-to-leading order (NNLO) QCD calculations being currently out of reach, it is useful to systematically consider at least some part of the higher order corrections to improve the theoretical precision. This can be achieved using resummation techniques for corrections originating from emission of soft gluons.  This type of emission happens in the presence of kinematical constraints, where the phase space available for emission of real gluons is restricted. As the kinematical limit is approached, the corrections are dominated by large logarithmic contributions with an argument of the logarithms directly related to the distance to the limit. The observable in question for which the predictions are obtained and the kinematics in which it is considered determines then the exact form of the logarithms.

Two popular approaches to perform soft gluon resummation are either direct calculations in QCD or an application of an effective field theory, in this case soft-collinear effective theory (SCET). Although the physics that is described is obviously the same, and the perturbative accuracy which can be reached is also formally the same, the two approaches differ at the technical level resulting in different treatment of subleading corrections beyond the formal accuracy. However, in practice these corrections can introduce non-negligible effects. Therefore it is valuable to perform calculations using both techniques, firstly as a completely independent check of the calculations and secondly as an indication of the size of subleading effects.

 As for any process for which production rates are small, also for the associated top-pair production with a heavy boson the first quantity which can be studied with higher precision is the total cross section. The higher order corrections receive then potentially large contributions from soft gluon emission in the threshold limit, i.e. the partonic center of mass energy $\shat$ approaches the energy needed to produce the final state with a given characteristics. 
For the process $pp\to \tth$ soft gluon resummation has been performed both using direct QCD~\cite{Kulesza:2015vda,Kulesza:2016vnq, Kulesza:2017ukk,Kulesza:2017jqv}  and SCET~\cite{Broggio:2015lya, Broggio:2016lfj} methods. While the next-to-leading-logarithmic (NLL) calculations~\cite{Kulesza:2015vda} were carried out in the absolute threshold limit, $\shat \to (2 m_t +m_H)^2$,  the later calculations~\cite{Kulesza:2016vnq, Kulesza:2017ukk,Kulesza:2017jqv,Broggio:2015lya, Broggio:2016lfj} opted for the invariant mass threshold limit, i.e.  $\shat \to Q^2$ with $Q^2 = (p_t +p_{\tb} + p_H)^2$. 
The resummed predictions are now known at the next-to-next-to-leading logarithmic (NNLL) accuracy in both approaches and are matched to the full NLO results to include all available information on the process. In the case of associated top-pair production with a heavy gauge boson, $W^+, W^-$ or $Z$, NLO+NNLL predictions obtained within the SCET framework are already available~\cite{Li:2014ula, Broggio:2016zgg,Broggio:2017kzi}, whereas for calculations in the direct QCD approach only NLO+NLL results have been communicated so far~\cite{Kulesza:2017hoc}. Here we close this gap and report on soft gluon resummation in this approach at the NLO+NNLL accuracy for the process  $pp\to \ttV$. Our calculations rely on the techniques described in~\cite{Kulesza:2017ukk}.  We present numerical results for the total cross sections and the invariant mass distributions as well as comment on the comparison between our results and those of~\cite{Li:2014ula,Broggio:2016zgg,Broggio:2017kzi}.

The paper is structured as follows: in section 2 we review the direct QCD approach applied before in the calculations for the process $pp \to \tth$ and now adapted to the $pp \to \ttV$ case. The numerical results are presented and discussed in section 3. The conclusions and the summary of our work  can be found in section 4.

\section{NNLL resummation in the triple invariant mass kinematics for $2 \to 3$ processes with two massive coloured particles in the final state}
\label{s:theory}

In the following, we use the direct QCD approach to resummation of soft gluon corrections at threshold in  Mellin space. In particular, we consider the threshold limit in the three particle invariant mass kinematics.
The Mellin transformation of the differential cross section $d\sigma_{pp \to \ttV}/ dQ^2$ is then performed w.r.t.\ the variable \mbox{$\rho = Q^2/S$}. Resummation provides a systematic treatment of logarithmic terms of the form $\als^n \left[ \log^m (1-z)/(1-z)\right]_+$, with $m\leq 2n-1$ and $z=Q^2/\shat$ which appear at all orders of the perturbative expansion in $\als$.
These logarithms then turn into logarithms of the Mellin moment $N$ in Mellin space, where the threshold limit $z \to 1$ corresponds to the limit $N \to \infty$.

We use the same framework as developed in~\cite{Kulesza:2017ukk} and in the following we consider a process $ ij \to k l V$, where $i,j$ are massless coloured partons, $k, l$ two massive quarks and $V$ a massive colour-singlet particle. The collective argument $\{m^2\}$ denotes all masses entering the calculations. The resummed partonic cross section up to NNLL accuracy can be written as: 
\begin{eqnarray}
\label{eq:res:fact_diag}
\frac{d\tilde \sigh^{{\rm (NNLL)}}_{ij\tosv kl V}}{dQ^2}&&\hspace{-0.9cm}(N, Q^2,\{m^2\}, \mur^2) = {\mathrm{Tr}}\left[ \mathbf{H}_R (Q^2, \{m^2\},\muf^2, \mur^2)\mathbf{\bar{U}}_R(N+1, Q^2,\{m^2\}, Q^2 ) \, \right. \nn
\\ 
&\times& \left. \mathbf{\tilde S}_R (N+1, Q^2, \{m^2\})\, \mathbf{{U}}_R(N+1, Q^2,\{m^2\}, Q^2 )  \right] \nonumber \\
&\times&\,\Delta^i(N+1, Q^2,\muf^2, \mur^2 ) \Delta^j(N+1, Q^2,\muf^2, \mur^2 ),
\end{eqnarray}
where $\mathbf{H}_R$, $\mathbf{\bar{U}}_R$, $\mathbf{U}_R$ and $\mathbf{\tilde S}_R$ are colour matrices and the trace is taken over colour space.
$\Delta^i$ and $\Delta^j$ represent the logarithmic contributions from the (soft-)collinear gluon emission from the initial state partons. They are universal functions, depending only on the emitting parton, and can be found for example in~\cite{Catani:1996yz,Bonciani:1998vc} up to NLL and in~\cite{Catani:2003zt} up to NNLL level.

The term $\mathbf{\bar{U}}_R\mathbf{\tilde S}_R\mathbf{{U}}_R$ originates from a solution of the renormalization group equation of the soft function, which describes the soft wide angle emission. It consists of the soft function evolution matrices $\mathbf{\bar{U}}_R$ and $\mathbf{{U}}_R$, as well as $\mathbf{\tilde S}_R$ which plays the role of a boundary condition of the renormalization group equation. In general the evolution matrices are given by path-ordered exponentials of a soft anomalous dimension matrix in the colour space. If the matrix in the argument of the path ordered exponential is diagonal, it reduces to a sum over simple exponentials. All colour matrices in eq.~(\ref{eq:res:fact_diag}) are expressed in the basis in which the one loop soft-anomalous dimension $\mathbf{\Gamma}^{(1)}_{ij\to klV}$, i.e. ${\cal O}(\als)$ coefficient in the perturbative expansion the soft anomalous dimension $\mathbf{\Gamma}_{ij\to klV}$
\begin{equation}
\mathbf{\Gamma}_{ij\to klV}\left(\als\right)= \left[ \left(\frac{\als}{\pi}\right) \mathbf{\Gamma}^{(1)}_{ij\to klV} +\left(\frac{\als}{\pi}\right)^2 \mathbf{\Gamma}^{(2)}_{ij\to klV}+\ldots   \right]
\end{equation}
is diagonal. The diagonalization is achieved by a colour basis transformation 
\begin{eqnarray}
\mathbf{\Gamma}^{(1)}_{R}  =  \mathbf{R}^{-1}\mathbf{\Gamma}^{(1)}_{ij\to klV} \mathbf{R},\\
\mathbf{\Gamma}^{(1)}_{R,IJ}=\lambda^{(1)}_I \delta_{IJ},
\end{eqnarray}
where $\mathbf{R}$ is the diagonalization matrix and $\lambda^{(1)}_I$ are the eigenvalues of $\mathbf{\Gamma}^{(1)}_{ij\to klV}$.  Correspondingly, all colour matrices in eq.~(\ref{eq:res:fact_diag}) carry a subscript $R$. $\mathbf{\Gamma}_{ij\to klV}$  has to be known up to $\mathbf{\Gamma}^{(1)}_{ij\to klV}$ to perform resummation with NLL accuracy and up to $\mathbf{\Gamma}^{(2)}_{ij\to klV}$ for NNLL. The one-loop soft anomalous dimension can be found in~\cite{Kulesza:2015vda} while the two loop soft anomalous dimension was derived in~\cite{Ferroglia:2009ep,Ferroglia:2009ii}.

In practice, we start with a description of the colour structure of the  $\ttW$ and $\ttZ$ processes in the s-channel colour basis, $\{ c_I^q\}$ and $\{c_I^g\}$ given by  their basis vectors:
$$c_{\bf 1}^q =  \delta^{\al_{i}\al_{j}} \delta^{\al_{k}\al_{l}}, \quad c_{\bf 8}^{q} = T^a_{\al_{i}\al_{j}} T^a_{\al_{k}\al_{l}},
$$
$$c_{\bf 1}^{g} = \delta^{a_i a_j} \, \delta^{\al_k
  \al_l}, \quad c_{\bf 8S}^{g}=  T^b _{\alpha_l \alpha_k} d^{b a_i a_j} ,\quad
c^{g}_{\bf 8A} = i T^b _{\alpha_l \alpha_k} f^{b a_i a_j} .
$$ 
Since at leading order $\ttW$ state is produced via $q\bar{q}'$ channel we only need the $\{ c_I^q\}$ basis for its description, whereas both  $\{ c_I^q\}$ and  $\{ c_I^g\}$ basis are needed to describe the $\ttZ$ production via 
the $q\bar{q}$ and the $gg$ channels.

The function $\mathbf{\tilde S}_R$ is obtained by transforming  the purely eikonal function $\mathbf{\tilde S}_{ij\to klV}$ ,
\beq
\mathbf{\tilde S}_{R}  =  \mathbf{R}^{\dagger}\mathbf{\tilde S}_{ij\to klV} \mathbf{R}\\
\eeq
with
\beq
\mathbf{\tilde S}_{ij\to klV}=\mathbf{\tilde S}^{\mathrm{(0)}}_{ij\to kl V} + \frac{\als}{\pi}\mathbf{\tilde S}^{\mathrm{(1)}}_{ij\to klV} + \dots
\label{eq:soft:nlo}
\eeq
calculated in the $s$-channel colour basis and 
\begin{equation}
\left(\mathbf{\tilde S}^{\mathrm{(0)}}_{ij\to klV}\right)_{IJ}=\mathrm{Tr}\left[c_I^\dagger c_J\right]. 
\end{equation}
NLL accuracy requires knowledge of $\mathbf{\tilde S}^{\mathrm{(0)}}_{ij\to klV}$ while NNLL accuracy requires $\mathbf{\tilde S}^{\mathrm{(1)}}_{ij\to klV}$.

Since the one-loop soft anomalous dimension is in general non diagonal in the triple invariant mass kinematics, in order to calculate the soft function evolution matrices up to NLL we use the diagonalization method of ~\cite{Kidonakis:1998nf}.
In this way the path ordered exponentials reduce to a sum over simple exponentials and   $\mathbf{\bar{U}}_R\mathbf{\tilde S}_R\mathbf{{U}}_R$ at NLL is  given by 
\begin{equation}
\mathbf{\bar{U}}_{R,IJ}\mathbf{\tilde S}_{R,JK}\mathbf{{U}}_{R,KL} = \mathbf{\tilde S}^{\mathrm{(0)}}_{R,IL} \, \exp\left[\frac{\log(1-2\lambda)}{2 \pi b_0} \left(\left( \lambda^{(1)} _{I}\right)^{*}+\lambda^{(1)} _{L}\right)\right]
\end{equation}
where $\lambda$ is defined as
\begin{equation}
\lambda = \als(\mur^2) b_0 \log N
\end{equation}
and 
$$b_0 = \frac{11 \CA-4 \nf \TR}{12\pi}.$$

Resummation up to NNLL encounters additional complexity due to the non-commutativity of $\mathbf{\Gamma}^{(1)}_{ij\to klV}$ and $\mathbf{\Gamma}^{(2)}_{ij\to klV}$. Therefore we employ the method detailed in~\cite{Buras:1979yt,Ahrens:2010zv} to recast the soft function evolution matrices into simple exponentials.
This results in
\begin{eqnarray}
\label{eq:UR}
\mathbf{U}_R(N,Q^2,\{m^2\},Q^2 )&=&\left(\mathbf{1}+\frac{\alpha_{\mathrm{s}}(\mur^2)}{\pi(1-2\lambda)}\mathbf{K}\right)
\left[e^{\,g_s(N)\overrightarrow{\lambda}^{(1)}} \right]_{D}
\left(\mathbf{1}-\frac{\alpha_{\mathrm{s}}(\mur^2)}{\pi}\, \mathbf{K}\right),\\
 \mathbf{\bar{U}}_R(N, Q^2,\{m^2\},Q^2 )&=&\left(\mathbf{1}-\frac{\alpha_{\mathrm{s}}(\mur^2)}{\pi}\, \mathbf{K}^{\dagger}\right) 
\left[e^{\,g_s(N)\left(\overrightarrow{\lambda}^{(1)}\right)^*} \right]_{D} \left(\mathbf{1}+\frac{\alpha_{\mathrm{s}}(\mur^2)}{\pi(1-2\lambda)}\mathbf{K}^{\dagger}\right),\nonumber \\
\label{eq:URb}
\end{eqnarray}
with
\begin{equation}
K_{IJ}=\delta_{IJ}{\lambda}^{\left(1\right)}_{I}\frac{b_1}{2b_0^2}-\frac{\left(\mathbf{\Gamma}^{(2)}_R\right)_{IJ}}{2\pi b_0+\lambda^{\left(1\right)}_{I}-\lambda^{\left(1\right)}_{J}}\, ,
\end{equation}
\begin{eqnarray}
g_s (N)= \frac{1}{2 \pi b_0} \left\{  \log(1-2\lambda) + \als(\mur^2) \left[   \frac{b_1}{b_0} \frac{ \log(1-2\lambda)}{ 1-2\lambda} - 2 \gamma_{\rm E} b_0  \frac{2\lambda}{1-2\lambda} \right. \right. \nonumber \\
\left. \left.  
+ \, b_0 \log\left( \frac{Q^2}{\mur^2} \right) \frac{2\lambda}{1-2\lambda} \right] \right\}
\label{eq:gsoft}
\end{eqnarray}
and
$$
b_1 = \frac{17 \CA^2-\nf \TR\left(10\CA+6\CF\right)}{24\pi^2}\,.
$$

The hard function $\mathbf{H}_R$ describes the hard scattering contributions and absorbs off-shell effects. It is independent of $N$ and given by a matrix in colour space, which is then also transformed into the $R$ colour space
\begin{equation}
\mathbf{H}_R = \mathbf{R}^{-1}\, \mathbf{H}_{ij\to klV} \, \left(\mathbf{R}^{-1}\right)^{\dagger} .
\end{equation}
The hard function matrix can be calculated perturbatively:
\beq
\mathbf{H}_{ ij\tosv kl V}= \mathbf{H}^{\mathrm{(0)}}_{ij\to klV} + \frac{\als}{\pi}\mathbf{H}^{\mathrm{(1)}}_{ij\to klV} + \dots
\label{eq:hard:nlo}
\eeq
In order to perform resummation up to NNLL knowledge of $\mathbf{H}^{\mathrm{(0)}}_{ij\to klV}$ and as well as $\mathbf{H}^{\mathrm{(1)}}_{ij\to klV}$ is required. While the leading contribution $\mathbf{H}^{\mathrm{(0)}}_{ij\to klV}$ can be calculated from the LO cross section of the process, the next order includes $N$-independent non-logarithmic contributions originating from virtual loops, real collinear terms and the evolution matrices $\mathbf{U}_R$ and $\mathbf{\bar{U}}_R$.
The virtual contributions are extracted from the {\tt PowHel} code \cite{Garzelli:2012bn,Garzelli:2011is,Kardos:2011na} and projected on the colour basis. Following the method proposed in~\cite{Beenakker:2011sf,Beenakker:2013mva} the real terms are derived from the infrared limit of the real corrections.

The resummed cross sections of different accuracy denoted by "res" in the following are matched with the full NLO cross section according to
\bear
\label{hires}
 \frac{d\sigma^{\rm (matched)}_{h_1 h_2 \tosv klV}}{dQ^2}(Q^2,\{m^2\},\muf^2, \mur^2) &=& 
\frac{d\sigma^{\rm (NLO)}_{h_1 h_2 \tosv kl V}}{dQ^2}(Q^2,\{m^2\},\muf^2, \mur^2) \\ \nn
&+&   \frac{d \sigma^{\rm
  (res-exp)}_
{h_1 h_2 \tosv kl V}}{dQ^2}(Q^2,\{m^2\},\muf^2, \mur^2) 
\eear
 with
\bear
\label{invmel}
&& \frac{d \sigma^{\rm
  (res-exp)}_{h_1 h_2 \tosv kl V}}{dQ^2} (Q^2,\{m^2\},\muf^2, \mur^2) \! =   \sum_{i,j}\,
\int_{\sf C}\,\frac{dN}{2\pi
  i} \; \rho^{-N} f^{(N+1)} _{i/h{_1}} (\muf^2) \, f^{(N+1)} _{j/h_{2}} (\muf^2) \nn \\ 
&& \! \times\! \left[ 
\frac{d \tilde\sigh^{\rm (res)}_{ij\tosv kl V}}{dQ^2} (N,Q^2,\{m^2\},\muf^2, \mur^2)
-  \frac{d \tilde\sigh^{\rm (res)}_{ij\tosv kl V}}{dQ^2} (N,Q^2,\{m^2\},\muf^2, \mur^2)
{ \left. \right|}_{\scriptscriptstyle({\rm NLO})}\, \! \right], 
\eear
where "res" = N(N)LL and "matched" = NLO + N(N)LL for the N(N)LL resummed results matched to NLO. The moments of the parton 
distribution functions $f_{i/h}(x, \muf^2)$ are 
defined in the standard way 
$$
f^{(N)}_{i/h} (\muf^2) \equiv \int_0^1 dx \, x^{N-1} f_{i/h}(x, \muf^2)\,,
$$
and $ d \sigh^{\rm
  (res)}_{ij\tosv kl V}/ dQ^2 \left. \right|_{\scriptscriptstyle({\rm NLO})}$ represents the perturbative expansion of the resummed cross section truncated at NLO. The inverse Mellin transform (\ref{invmel}) is evaluated numerically using 
a contour ${\sf C}$ in the complex-$N$ space according to the ``Minimal Prescription'' 
method proposed in Ref.~\cite{Catani:1996yz}.

Apart from the NLO+NLL and NLO+NNLL results we also calculate the NLL result improved by including contributions of order $\cal{O}(\als)$ terms in $\mathbf{H}_{R}$ and $\mathbf{\tilde S}_R$ and matched to NLO which we refer to as NLO+NLL w ${\cal C}$ . The resummed partonic cross section at this accuracy is given by:
\begin{eqnarray}
\frac{d \tilde\sigh^{{\rm (NLL\ w\ {\cal C})}}_{ij\tosv kl V}}{dQ^2}&&\hspace{-0.9cm}(N,Q^2,\{m^2\},\muf^2, \mur^2) =  \mathbf{H}_{R,IJ}(Q^2, \{m^2\},\muf^2, \mur^2) \, \mathbf{\tilde S}_{R,JI}(Q^2,\{m^2\})\nn \\
  &\times&
 \Delta^i(N+1, Q^2,\muf^2, \mur^2) \Delta^j(N+1, Q^2,\muf^2, \mur^2) \nonumber \\
&\times& \, \exp\left[\frac{\log(1-2\lambda)}{2 \pi b_0}
\left(\left( \lambda^{(1)} _{J}\right)^{*}+\lambda^{(1)} _{I}\right)\right] ,
\label{eq:res:fact_diag_NLLwC}
\end{eqnarray}
where
\beq
\mathbf{H}_{R}\, \mathbf{\tilde S}_{R} = \mathbf{H}^{\mathrm{(0)}}_{R} \mathbf{\tilde S}^{\mathrm{(0)}}_{R} + \frac{\als}{\pi}\left[\mathbf{H}^{\mathrm{(1)}}_{R} \mathbf{\tilde S}^{\mathrm{(0)}}_{R}+ \mathbf{H}^{\mathrm{(0)}}_{R} \mathbf{\tilde S}^{\mathrm{(1)}}_{R} \right] . \nn
\eeq

\section{Numerical results for the $pp \to \ttV$ processes at NLO+NNLL accuracy}
\label{s:results}
In this section we present our resummed results with different levels of precision e.g.\ NLL, NLL w ${\cal C}$ and NNLL matched to NLO. They include distributions differential in $Q$ as well as total cross sections which were calculated by integrating over $Q$. 
The resummed results were obtained with two independently developed in-house codes, while the NLO cross sections were calculated with {\tt MadGraph5\_aMC@NLO}~\cite{Alwall:2014hca} for differential distributions and total cross sections, and with {\tt PowHel}~\cite{Garzelli:2012bn,Garzelli:2011is,Kardos:2011na} for NLO total cross sections without the contributions from $qg$ channels. In the calculations we use the  PDF4LHC15\_30 parton distribution function (pdf) set~\cite{Butterworth:2015oua, Dulat:2015mca, Harland-Lang:2014zoa, Ball:2014uwa, Gao:2013bia, Carrazza:2015aoa}  and input parameters according to the Higgs Cross Section Working Group (HXSWG)  recommendations~\cite{hxswg}, i.e. $m_H=125$ GeV, $m_t=172.5$ GeV, $m_W=80.385$ GeV, $m_Z=91.188$ GeV, $G_F=1.1663787\cdot10^{-5}\, \text{GeV}^{-2}$. This is the same choice as the one made in the HXSWG Yellow Report 4~\cite{deFlorian:2016spz}, so that we can reproduce the NLO values of the $\ttV$ cross sections listed there and compare our new resummed predictions to them.  In accordance with the Yellow Report setup, in the calculations of the $\ttW$ cross sections the CKM matrix is taken diagonal. NLO pdf sets are used for NLO, NLL matched to NLO and NLL w ${\cal C}$ matched to NLO results, while NNLO pdf sets for NNLL matched to NLO. The pdf error is only calculated for the NLO cross sections for technical simplicity, since resummation will influence the pdf error only minimally. 
Two different choices for the central factorisation and renormalisation scales are used for most of the results throughout the section. The first choice is $\mu_0=\mufo=\muro=Q$ which is the natural scale for the threshold and kinematics in the resummation, while the second scale is $\mu_0=\mufo=\muro=M/2$, which is often used in fixed order calculations. 
The scale error is calculated using the seven point method by taking the maximum and minimum of the scale variations $(\muf/\mu_{0}, \mur/\mu_{0}) = (0.5,0.5), (0.5,1), (1,0.5), (1,1), (1,2), (2,1), (2,2)$.

As mentioned in the previous section the hard function includes virtual loop corrections which were extracted from {\tt PowHel}~\cite{Garzelli:2012bn,Garzelli:2011is,Kardos:2011na}. Analytical relations between the basis colour configurations of the colour flow basis used in {\tt PowHel} and the basis vectors of the s-channel colour basis allow us to extract the full matrix $\mathbf{H}_{R,IJ}$.
The colour summed results were then compared to the standalone MadLoop implementation from {\tt MadGraph5\_aMC@NLO}~\cite{Alwall:2014hca}.

\subsection{Total cross sections}
\label{s:totalxs}
\begin{figure}[h!]
\centering
\includegraphics[width=0.45\textwidth]{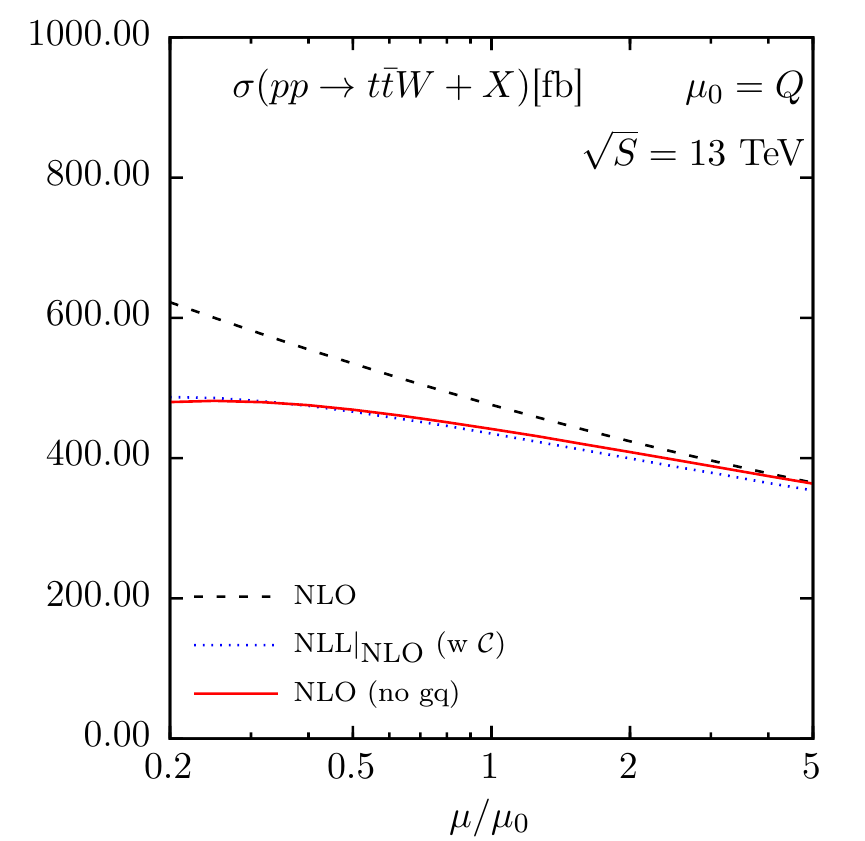}
\includegraphics[width=0.45\textwidth]{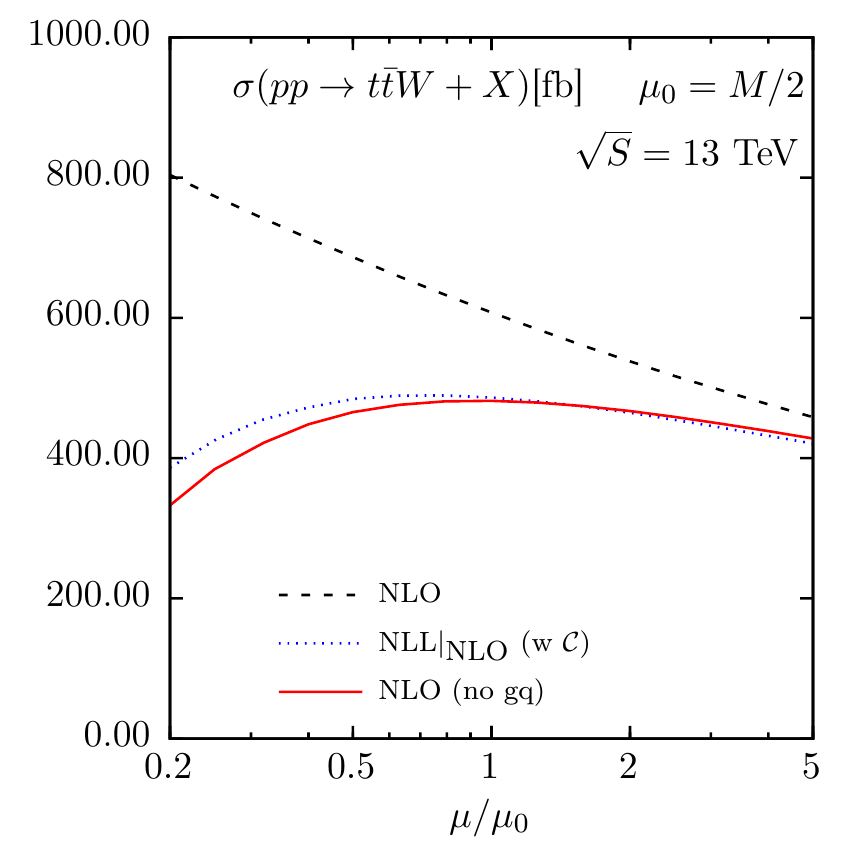}
\caption{Comparison between the resummed expression expanded up to NLO accuracy in $\als$, the full NLO result and the NLO result without the $qg$ channel for the $\ttW$ production.} 
\label{f:NLL_expansion_vs_NLO_ttW}
\end{figure}

\begin{figure}[h!]
\centering
\includegraphics[width=0.45\textwidth]{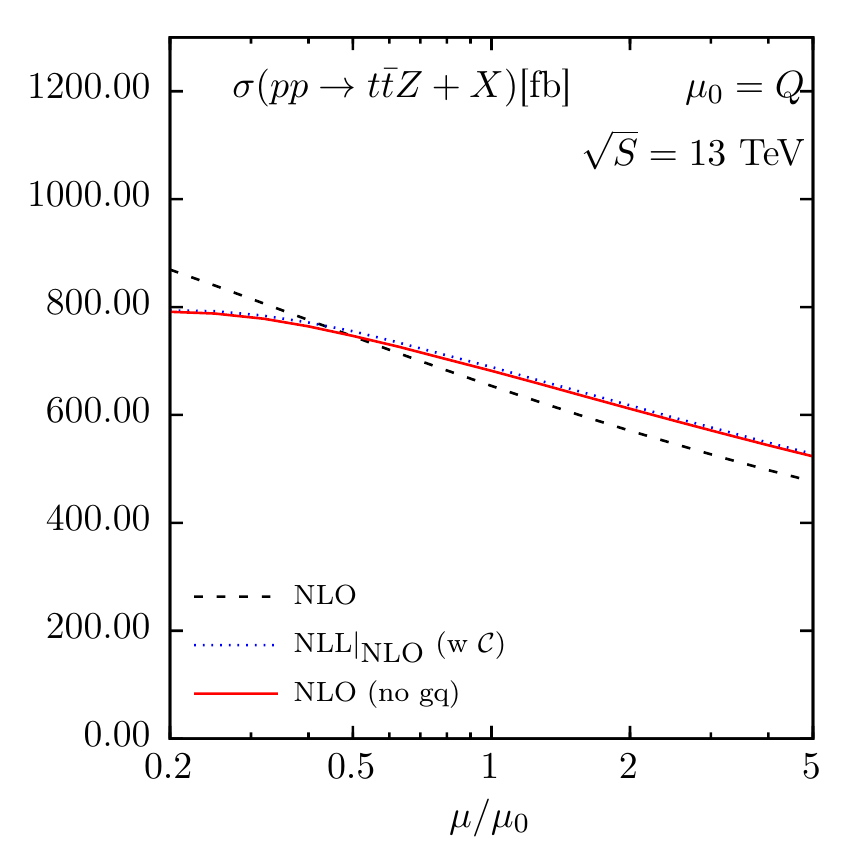}
\includegraphics[width=0.45\textwidth]{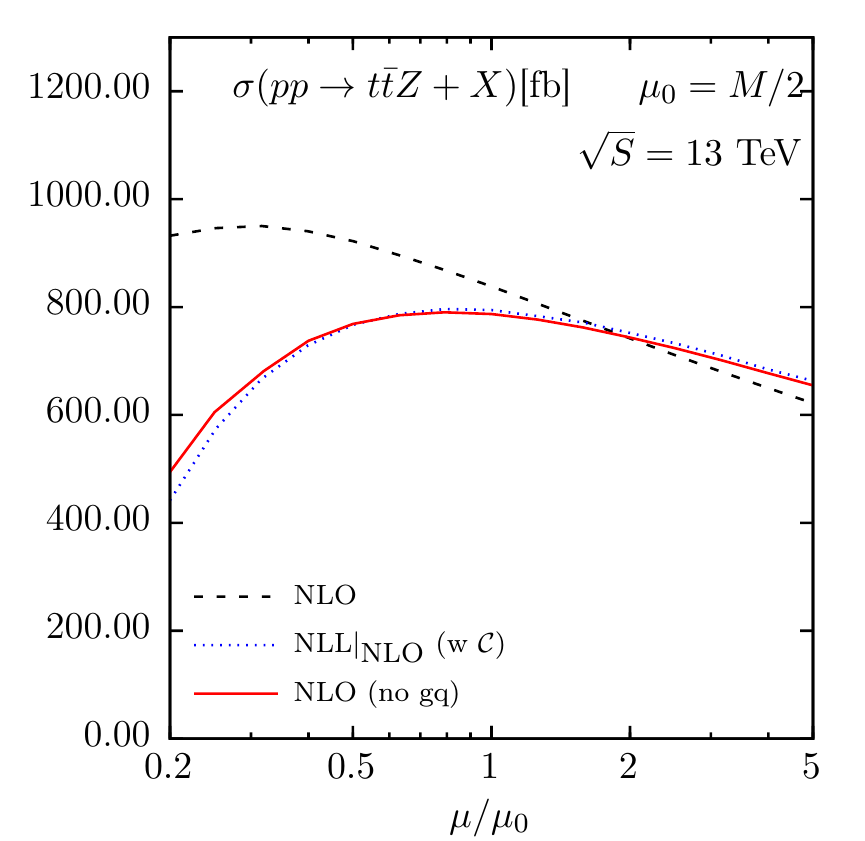}
\caption{Comparison between the resummed expression expanded up to NLO accuracy in $\als$, the full NLO result and the NLO result without the $qg$ channel for the $\ttZ$ production.} 
\label{f:NLL_expansion_vs_NLO_ttZ}
\end{figure}

At first we compare the total cross section of the full NLO calculation with our resummed result expanded in $\als$ up to NLO to analyse how well the full NLO cross section can be reproduced.
The relatively large significance of the $qg$ channel especially for the scale uncertainties was shown first in~\cite{Li:2014ula} for $\ttW$ and later in~\cite{Broggio:2016zgg,Broggio:2017kzi} for $\ttW$ and $\ttZ$.
Since the $qg$ channel first appears at NLO, no resummation is performed for this channel. Therefore one has to compare the NLO cross section without the $qg$ channel with the resummed result expanded in $\als$ to judge the quality of the approximation provided by the expansion. 
The resummed cross sections matched to NLO include the $qg$ channel through the matching to the full NLO calculations. In figure \ref{f:NLL_expansion_vs_NLO_ttW} we compare of the full NLO cross section, the NLO cross section without the $qg$ channel and the expansion of the resummed cross section as a function of ${\mu}/{\mu_0}={\muf}/{\mufo}={\mur}/{\muro}$ for $\ttW$ at $\sqrt S=13$ TeV for the two different scale choices $\mu_0=Q$ and $\mu_0=M/2$.
The corresponding comparison for $\ttZ$ is shown in figure~\ref{f:NLL_expansion_vs_NLO_ttZ}.

In all cases the NLO cross section without the $qg$ channel is much better approximated by the expansion of the resummed cross section than the full NLO result. Because of the good agreement between NLO without the $qg$ channel and the expanded result we conclude that the resummation includes a big part of the higher order corrections for the production channels present at LO.

\begin{table}
	\begin{center}
\renewcommand{\arraystretch}{1.5}
		\begin{tabular}{|c c c c c c c|}
			\hline
			process & $\mu_0$ & NLO{[}fb{]} & {NLO+NLL}{[}fb{]} & {NLO+NLL w $\cal C$}{[}fb{]} & {NLO+NNLL}{[}fb{]} & $K_{\text{NNLL}}$ \tabularnewline
			\hline	 
			$t \bar t W^{+}$ & $Q$ & $323_{-10.8\%}^{+12.2\%}$ & $325_{-10.4\%}^{+11.8\%}$  & $336_{-9.2\%}^{+9.8\%}$  & $342_{-8.6\%}^{+8.9\%}$ & $1.06$  \tabularnewline
			& $Q/2$ & $363_{-10.9\%}^{+12.1\%}$ & $364_{-10.6\%}^{+11.9\%}$  & $368_{-9.1\%}^{+10.4\%}$  & $371_{-8.7\%}^{+9.7\%}$ & $1.02$ \tabularnewline
			& $M/2$ & $413_{-11.4\%}^{+12.7\%}$ & $414_{-11.3\%}^{+13.1\%}$  & $413_{-10.0\%}^{+13.0\%}$  & $415_{-9.6\%}^{+12.9\%}$ & $1.01$ \tabularnewline	
			\hline 
			$t \bar t W^{-}$  & $Q$ & $163_{-10.9\%}^{+12.5\%}$ & $165_{-10.4\%}^{+12\%}$  & $171_{-9.2\%}^{+9.9\%}$ &  $176_{-8.6\%}^{+8.8\%}$ & $1.08$ \tabularnewline
			& $Q/2$ & $184_{-11.1\%}^{+12.4\%}$ & $185_{-10.7\%}^{+12.1\%}$  & $187_{-9.1\%}^{+10.4\%}$  & $191_{-8.7\%}^{+9.6\%}$ & $1.04$ \tabularnewline 
			& $M/2$ & $208_{-11.6\%}^{+13.4\%}$ & $209_{-11.4\%}^{+13.8\%}$  & $209_{-9.9\%}^{+13.5\%}$  & $212_{-9.5\%}^{+13.2\%}$ & $1.02$ \tabularnewline
			\hline 
			$t \bar t Z$ & $Q$ & $659_{-12.7\%}^{+14.1\%}$ & $696_{-10.2\%}^{+11.7\%}$  & $795_{-9.8\%}^{+10.8\%}$ &  $848_{-8.3\%}^{+8.3\%}$ & $1.29$ \tabularnewline
			& $Q/2$ & $752_{-12.4\%}^{+12.7\%}$ & $770_{-9.6\%}^{+10.8\%}$  & $825_{-8.9\%}^{+8.9\%}$  & $856_{-7.9\%}^{+7.2\%}$ & $1.14$ \tabularnewline 
			& $M/2$ & $843_{-11.3\%}^{+9.7\%}$ & $850_{-9.8\%}^{+11.5\%}$  & $861_{-7.9\%}^{+7.3\%}$  & $875_{-7.7\%}^{+7.0\%}$ & $1.04$ \tabularnewline
			\hline 
		\end{tabular}
	\end{center}
\caption{Total cross section predictions for $pp \to t \bar t W^{+}/W^{-}/Z$ at $\sqrt S=13$ TeV and different central scale choices. The listed error is the theoretical error due to scale variation calculated using the seven point method.}
\label{t:totalxsec_13TeV}
\end{table}

\begin{table}
	\begin{center}
\renewcommand{\arraystretch}{1.5}
		\begin{tabular}{|c c c c c c c|}
			\hline
			process & $\mu_0$ & NLO{[}fb{]} & {NLO+NLL}{[}fb{]} & {NLO+NLL w $\cal C$}{[}fb{]} & {NLO+NNLL}{[}fb{]} & $K_{\text{NNLL}}$ \tabularnewline
			\hline	 
			$t \bar t W^{+}$ & $Q$ & $370_{-10.8\%}^{+12.2\%}$ & $372_{-10.4\%}^{+11.9\%}$  & $384_{-9.2\%}^{+10.0\%}$  & $391_{-8.7\%}^{+9.0\%}$ & $1.06$  \tabularnewline
			& $Q/2$ & $415_{-10.9\%}^{+12.2\%}$ & $416_{-10.6\%}^{+12.1\%}$  & $421_{-9.2\%}^{+10.6\%}$  & $425_{-8.9\%}^{+10.0\%}$ & $1.02$ \tabularnewline
			& $M/2$ & $474_{-11.5\%}^{+13.1\%}$ & $476_{-11.5\%}^{+13.6\%}$  & $475_{-10.2\%}^{+13.5\%}$  & $478_{-9.8\%}^{+13.4\%}$ & $1.01$ \tabularnewline	
			\hline 
			$t \bar t W^{-}$  & $Q$ & $191_{-10.9\%}^{+12.6\%}$ & $192_{-10.4\%}^{+12.1\%}$  & $199_{-9.2\%}^{+10.1\%}$ &  $205_{-8.6\%}^{+9.0\%}$ & $1.07$ \tabularnewline
			& $Q/2$ & $215_{-11.2\%}^{+12.6\%}$ & $216_{-10.8\%}^{+12.4\%}$  & $219_{-9.3\%}^{+10.7\%}$  & $222_{-8.9\%}^{+9.9\%}$ & $1.04$ \tabularnewline 
			& $M/2$ & $245_{-11.7\%}^{+14.2\%}$ & $245_{-11.5\%}^{+14.5\%}$  & $245_{-10.1\%}^{+14.3\%}$  & $249_{-9.8\%}^{+13.7\%}$ & $1.02$ \tabularnewline
			\hline 
			$t \bar t Z$  & $Q$ & $799_{-12.6\%}^{+13.9\%}$ & $843_{-10.2\%}^{+11.6\%}$  & $963_{-9.9\%}^{+10.7\%}$ &  $1028_{-8.4\%}^{+8.3\%}$ & $1.29$ \tabularnewline
			& $Q/2$ & $910_{-12.2\%}^{+12.6\%}$ & $931_{-9.5\%}^{+10.8\%}$  & $998_{-8.9\%}^{+9.0\%}$  & $1036_{-8.0\%}^{+7.3\%}$ & $1.14$ \tabularnewline 
			& $M/2$ & $1023_{-11.3\%}^{+9.7\%}$ & $1031_{-9.9\%}^{+11.6\%}$  & $1042_{-8.1\%}^{+7.4\%}$  & $1062_{-7.9\%}^{+7.0\%}$ & $1.04$ \tabularnewline
			\hline 
		\end{tabular}
	\end{center}
\caption{Total cross section predictions for $pp \to t \bar t W^{+}/W^{-}/Z$ at $\sqrt S=14$ TeV and different central scale choices. The listed error is the theoretical error due to scale variation calculated using the seven point method.}
\label{t:totalxsec_14TeV}
\end{table}

\begin{figure}[h]
\centering
\includegraphics[width=0.45\textwidth]{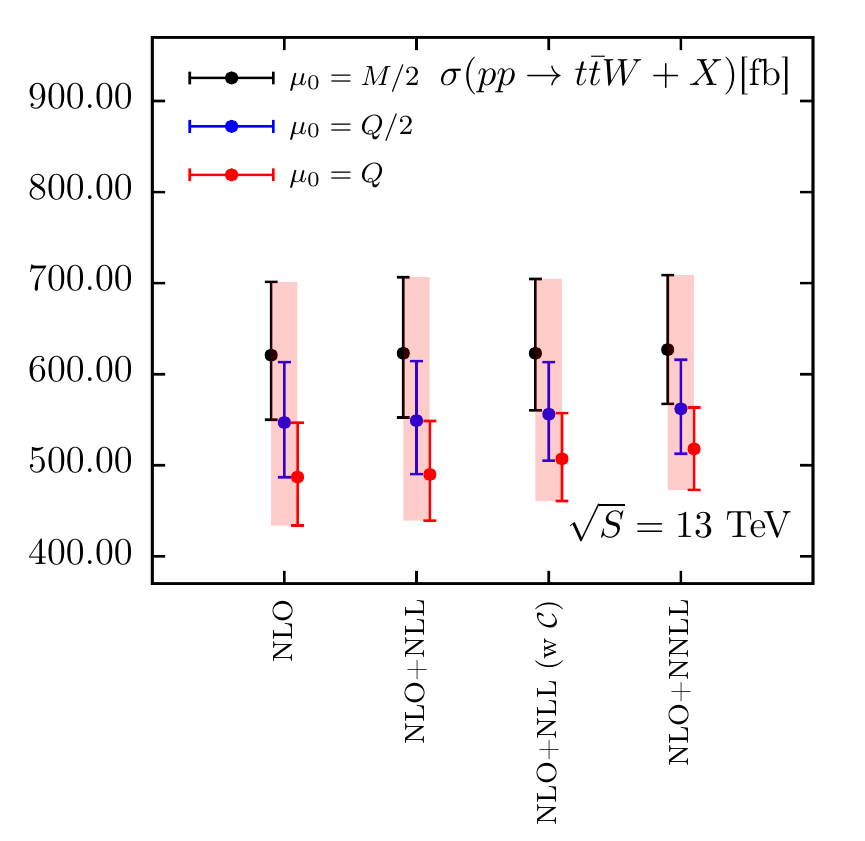}
\includegraphics[width=0.45\textwidth]{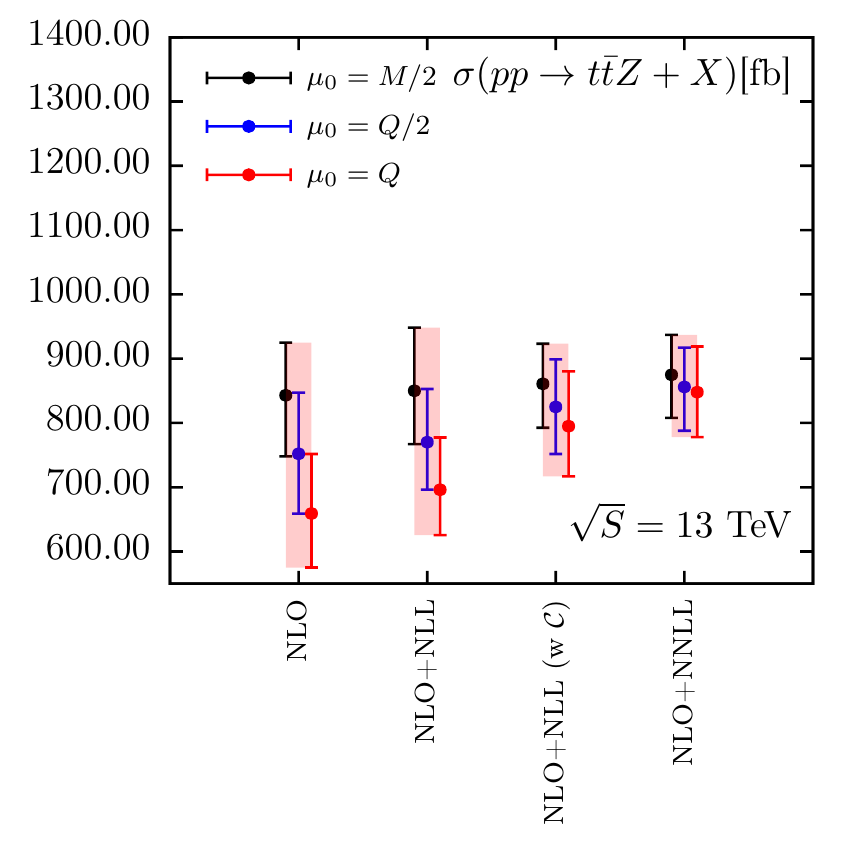}
\caption{Graphical illustration of results presented in table~\ref{t:totalxsec_13TeV}.} 
\label{f:totalxsec_13TeV}
\end{figure}

\begin{figure}[h]
\centering
\includegraphics[width=0.45\textwidth]{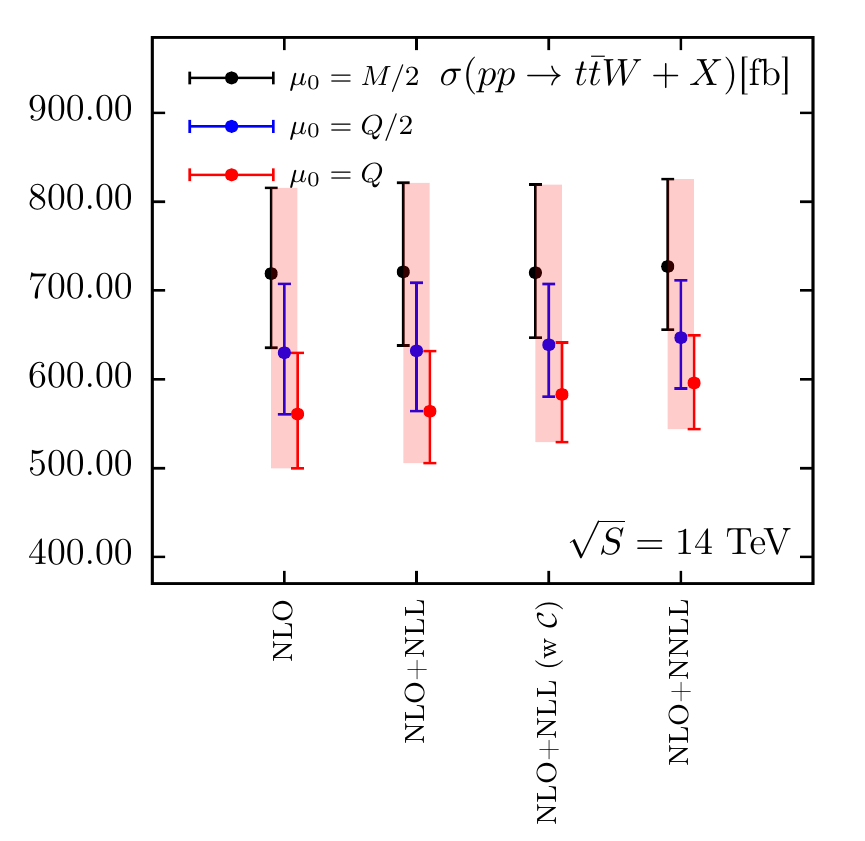}
\includegraphics[width=0.45\textwidth]{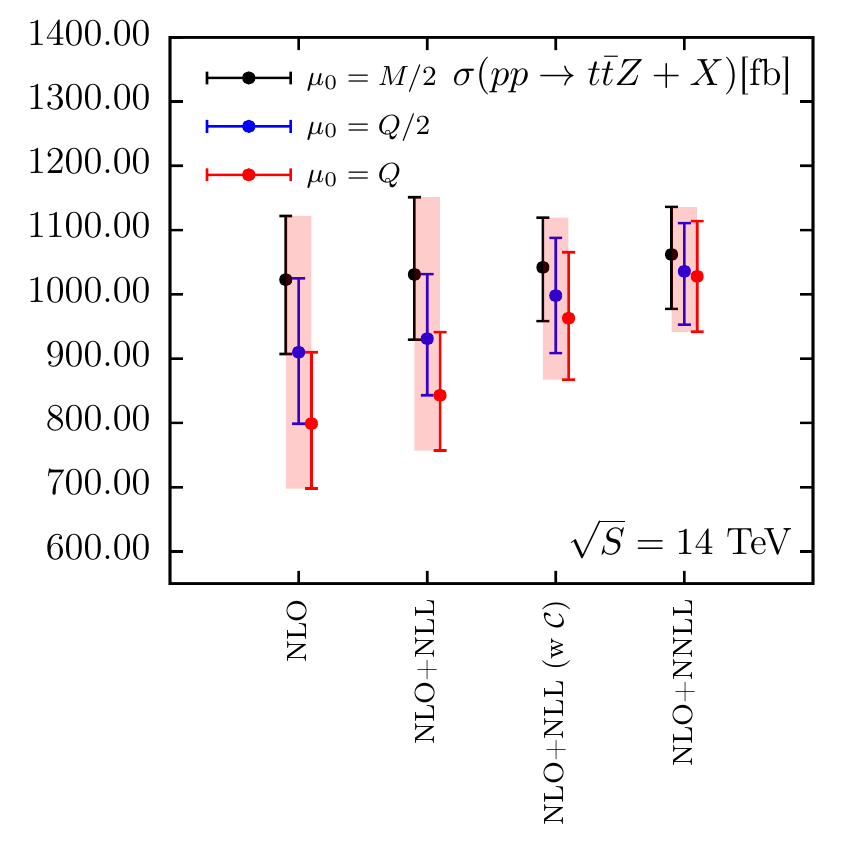}
\caption{Graphical illustration of results presented in table~\ref{t:totalxsec_14TeV}.} 
\label{f:totalxsec_14TeV}
\end{figure}

Predictions for the total cross section for $\sqrt S=13$ TeV and $\sqrt S=14$ TeV are shown in tables \ref{t:totalxsec_13TeV} and \ref{t:totalxsec_14TeV}. These results are visualised in figures~\ref{f:totalxsec_13TeV} and~\ref{f:totalxsec_14TeV}. They show the predictions with their scale uncertainties for the three central scales $\mu_0=M/2$, $\mu_0=Q$ and $\mu_0=Q/2$ as an `in-between scale choice'. The NLO values listed here fully agree with the NLO QCD cross sections published in the HXSWG Yellow Report 4~\cite{deFlorian:2016spz} within statistical Monte Carlo errors. Although the NLO results for various scale choices span quite a large range of values, the NLO+NNLL results are considerably closer, indicating the importance of resummed calculations. In general, the range of values spanned by the results decreases as the precision of the calculations increases. Another manifestation of the same effect originating from soft gluon corrections is the decrease in the scale uncertainties calculated for each specific scale choice which is also progressing with increasing precision of the theoretical predictions. This trend is much stronger for $\ttZ$ production than for $\ttW$ due to the $gg$ channel contributing to the LO and, correspondingly, to the resummed cross section. As the gluon radiate more than quarks, resummation has more relevance for the $gg$ production channel than for $q \bar q$ or $q \bar q^{'}$ channels. Correspondingly, we see a decrease in the $\ttZ$ scale uncertainty of about 30--40\% when increasing the precision from NLO to NLO+NNLL. 
The $\ttW$ cross section scale uncertainty is reduced by 20--30\% with the exception of the upwards uncertainty for $\mu=M/2$ which does not receive any significant improvement. 

As already noted, the NLO+NNLL predictions with the central scale varied between $M/2$, $Q/2$ and $Q$ are closer in value than the corresponding NLO predictions. The NLO+NNLL $\ttZ$ results are particularly stable w.r.t.\ scale variation. Correspondingly, the NNLL $K$-factors, ranging from 1.04 to 1.29, cf. tables~\ref{t:totalxsec_13TeV} and ~\ref{t:totalxsec_14TeV}, have to compensate for the scale dependence of the NLO results. Due to limited  corrections from resummation for the quark-initiated $\ttW$ process the NNLL $K$-factors are smaller, ranging from 1.01 to 1.07. The scale dependence of the NLO is strongly influenced by the scale dependence of the $qg$ channel (see e.g.~\cite{Campbell:2012dh}) which is formally subleading and not resummed here. Therefore resummation for the $q\bar q'$ channel does not fully compensate the scale dependence of the NLO and leads only to moderate improvements at NLO+NNLL.

Note that it is also possible to obtain soft gluon approximation of the NNLO corrections by expanding the resummed cross section. We performed such studies for the $\tth$ production at the LHC~\cite{Kulesza:2017ukk}, where we added this approximation to the full NLO result, resulting in the NNLO$_{\rm Approx.}$ predictions. We found them to be fully consistent with the resummed NLO+NNLL cross sections. For the $\ttV$ processes, the approximate NNLO predictions were already presented in~\cite{Broggio:2016zgg, Broggio:2017kzi}. Since here we are interested in the resummed results, we refer the readers interested in NNLO$_{\rm Approx.}$ predictions to these publications. 

The observed improvement in the stability of the predictions w.r.t. scale variation at NLO+NNLL for the  $\ttZ$ process is akin to the improvement for the $\tth$ process~\cite{Kulesza:2017ukk}. Similarly, we are encouraged to combine the predictions for our three representative scale choices according to the envelope method proposed by the HXSWG~\cite{Dittmaier:2011ti}. This way we can obtain theoretical predictions with the most conservative estimate of the scale error. The corresponding result for the $\ttZ$ production at 13 TeV is:
\begin{equation}
\sigma^{\ttZ}_{\rm NLO+NNLL}=863 ^{+8.5 \% +3.2 \%}_{-9.9 \% -3.2 \%} \ {\rm fb},
\label{eq:totalxsec_ttZ_combined}
\end{equation}
and at 14 TeV
$$
\sigma^{\ttZ}_{\rm NLO+NNLL}=1045 ^{+8.8 \% +3.1 \%}_{-9.9 \% -3.1 \%} \ {\rm fb}.
$$
The first uncertainty originates from the scale variation and is calculated using the envelope method, whereas  the second one is the pdf+$\als$ uncertainty. These values are in a very good agreement with the NLO results obtained for the scale choice  $\mu_0 = \mufo=\muro= M/2$,  justifying this common choice to obtain theory predictions for this process.

The same treatment can be applied to the $\ttWp$ and $\ttWm$ production resulting in
\begin{equation}
\sigma^{\ttW^{+}}_{\rm NLO+NNLL}=374 ^{+25.3 \% +3.2 \%}_{-16.4 \% -3.2 \%} \ {\rm fb},
\label{eq:totalxsec_ttWp_combined}
\end{equation}
\begin{equation}
\sigma^{\ttW^{-}}_{\rm NLO+NNLL}=192 ^{+25.2 \% +3.7 \%}_{-16.1 \% -3.7 \%} \ {\rm fb},
\label{eq:totalxsec_ttWm_combined}
\end{equation}
for $\sqrt{S}=13$~TeV, and at $\sqrt{S}=14$~TeV
$$
\sigma^{\ttW^{+}}_{\rm NLO+NNLL}=429 ^{+26.4 \% +3.2 \%}_{-16.7\% -3.2 \%} \ {\rm fb},
$$
$$
\sigma^{\ttW^{-}}_{\rm NLO+NNLL}=224 ^{+26.4 \% +3.6 \%}_{-16.4 \% -3.6 \%} \ {\rm fb},
$$
where again the first uncertainty originates from the scale variation and the second is the pdf+$\als$ uncertainty.
Due to a worse agreement between cross section predictions for the different central scale choices this treatment leads to a larger uncertainty than the uncertainty for the common choice of $\mu=M/2$. 

\begin{figure}[h!]
\centering
\includegraphics[width=0.45\textwidth]{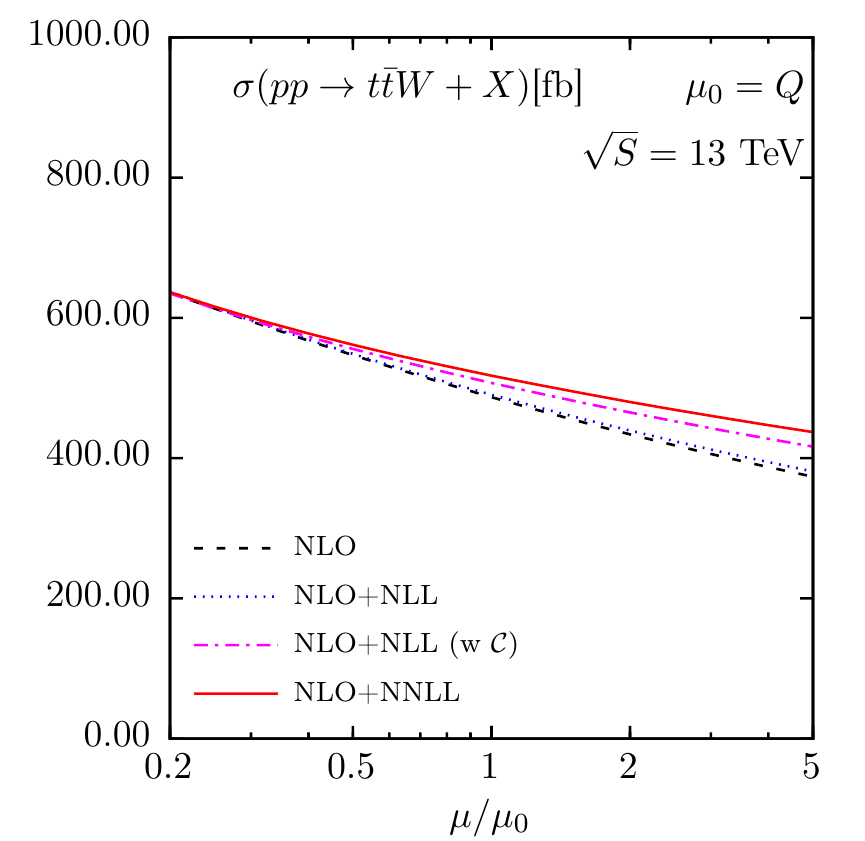}
\includegraphics[width=0.45\textwidth]{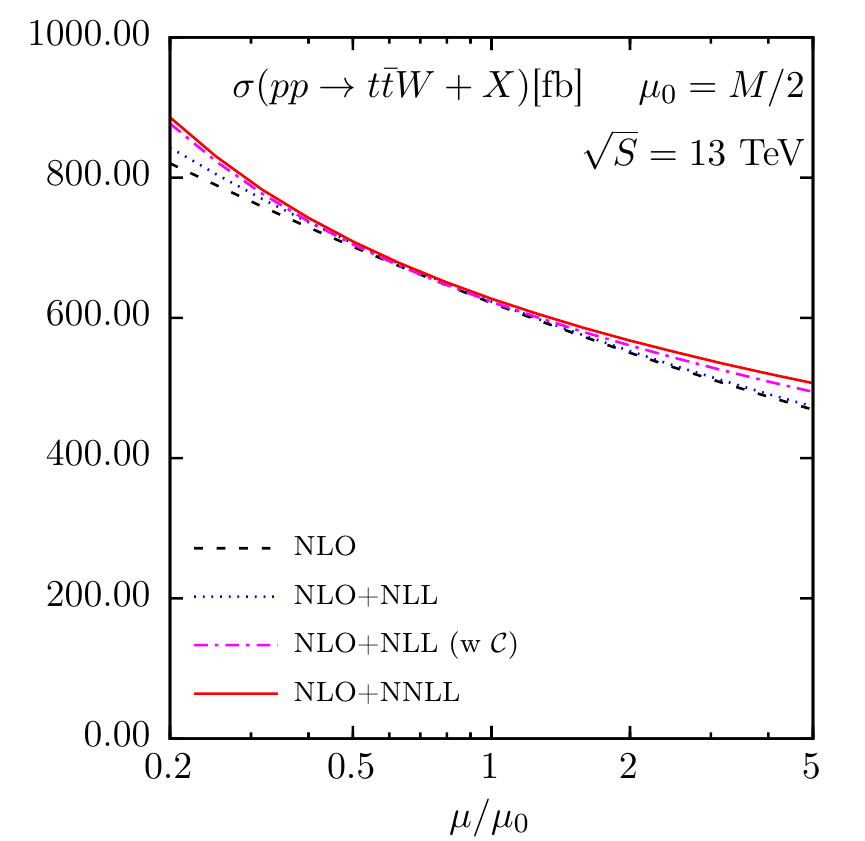}
\caption{Scale dependence of the total cross section for the process $pp\to \ttW$ at the LHC with $\sqrt S=13$ TeV. Results are shown for the choice $\mu=\muf=\mur$ and two central scale values $\mu_0=Q$ (left plot) and $\mu_0=M/2$ (right plot).} 
\label{f:scaledependence_ttW}
\end{figure}

\begin{figure}[h!]
\centering
\includegraphics[width=0.45\textwidth]{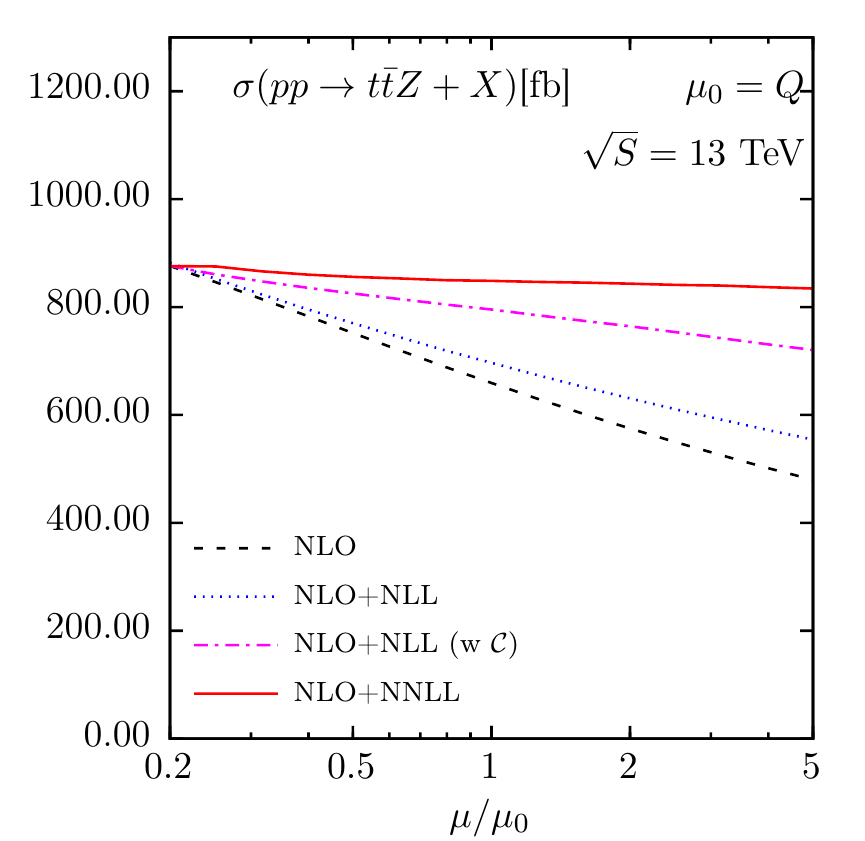}
\includegraphics[width=0.45\textwidth]{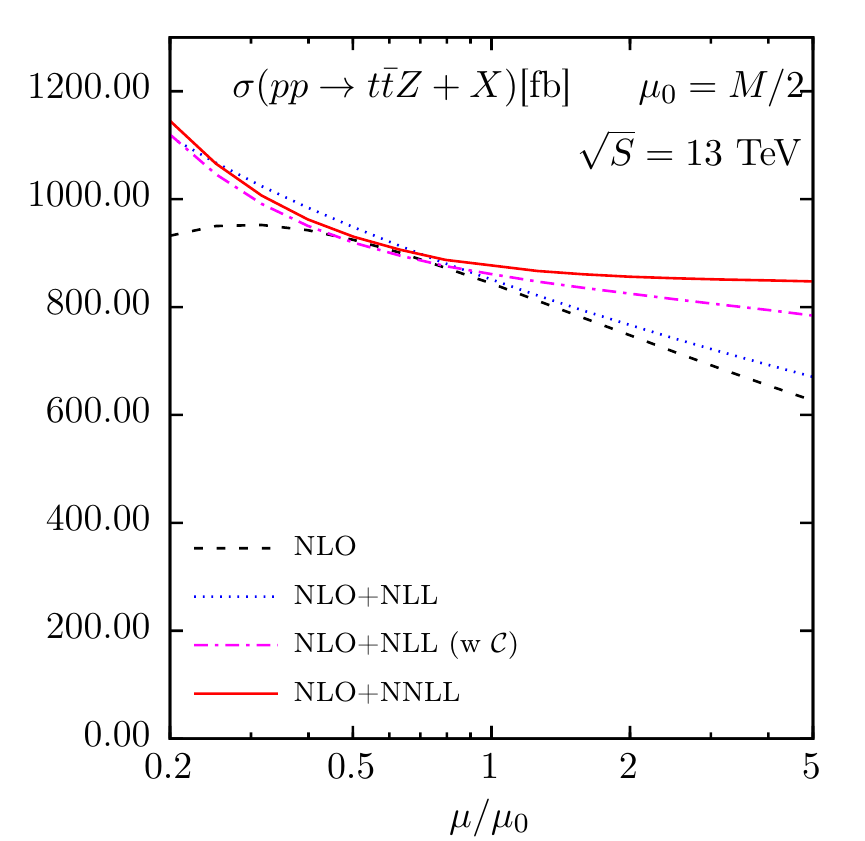}
\caption{Scale dependence of the total cross section for the process $pp\to \ttZ$ at the LHC with $\sqrt S=13$ TeV. Results are shown for the choice $\mu=\muf=\mur$ and two central scale values $\mu_0=Q$ (left plot) and $\mu_0=M/2$ (right plot).} 
\label{f:scaledependence_ttZ}
\end{figure}

\begin{figure}[t!]
\centering
\includegraphics[width=0.45\textwidth]{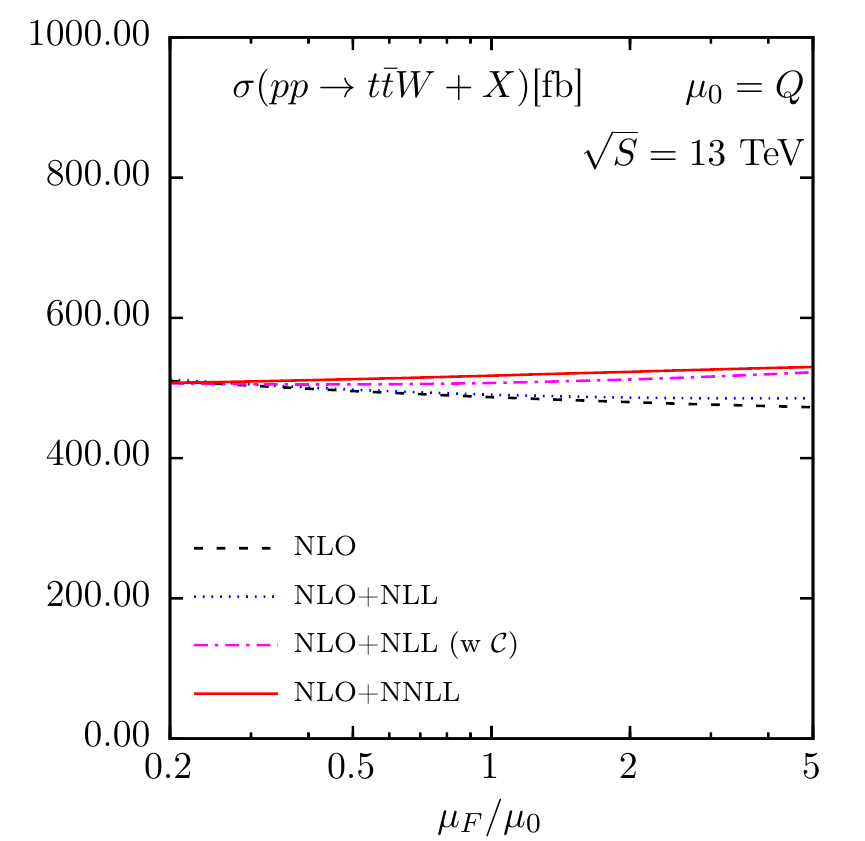}
\includegraphics[width=0.45\textwidth]{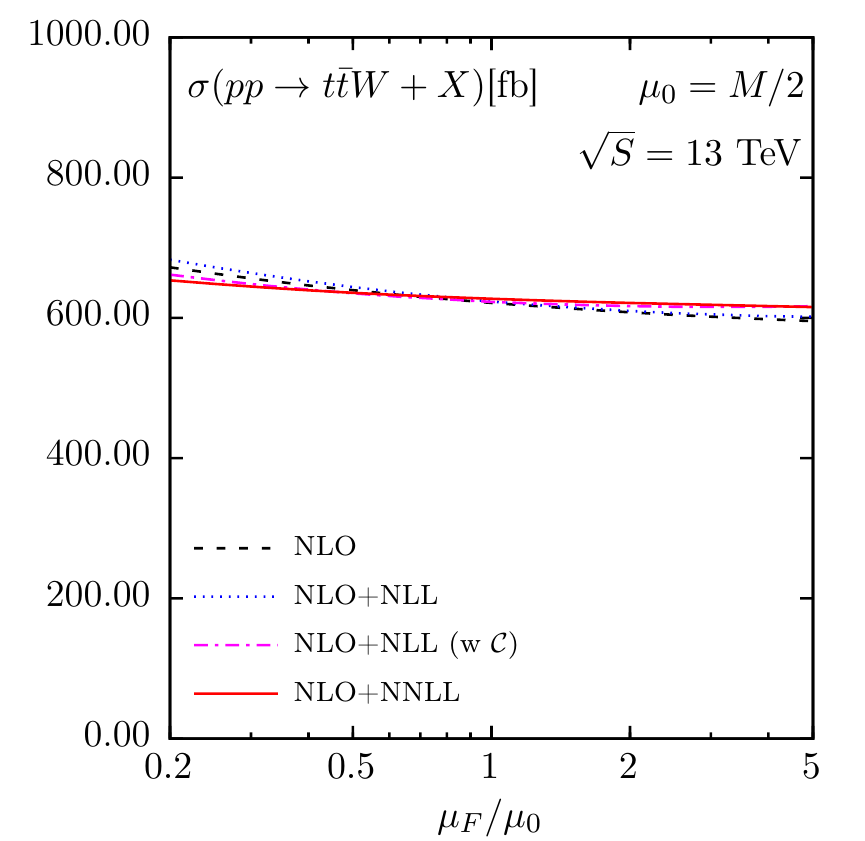}
\caption{Factorisation scale dependence of the total cross section for the process $pp\to \ttW$ at the LHC with $\sqrt S=13$ TeV and $\mur=\mu_{R,0}$ kept fixed. Results are shown for two central scale values $\mu_0=\mufo=\muro=Q$ (left plot) and $\mu_0=\mufo=\muro=M/2$ (right plot).} 
\label{f:mufdep_ttW}
\end{figure}

\begin{figure}[h!]
\centering
\includegraphics[width=0.45\textwidth]{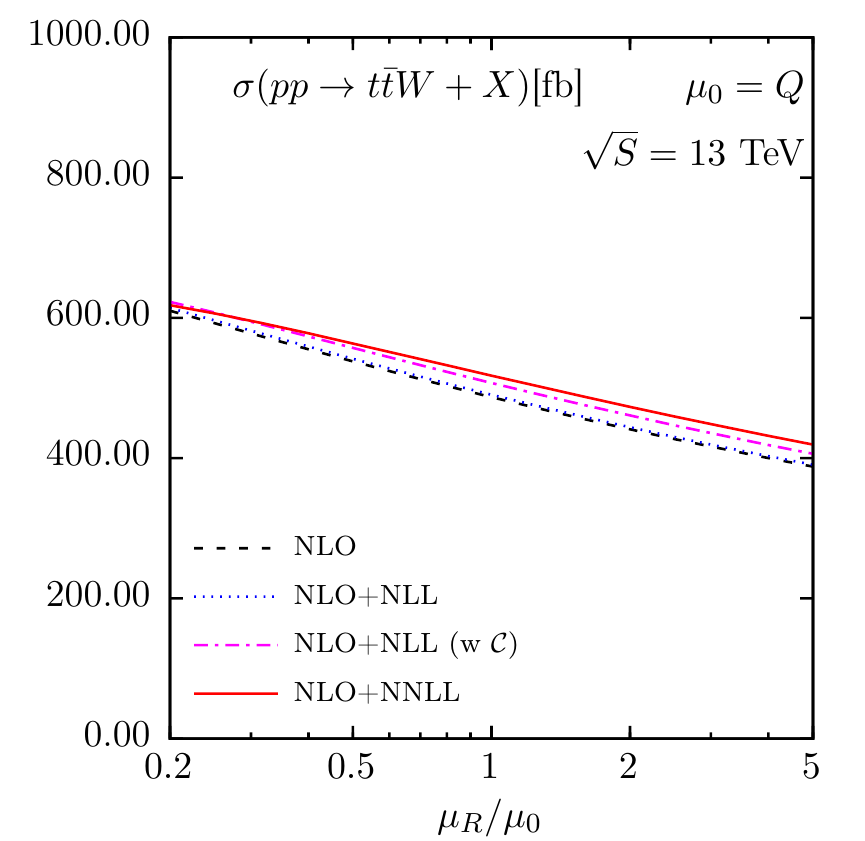}
\includegraphics[width=0.45\textwidth]{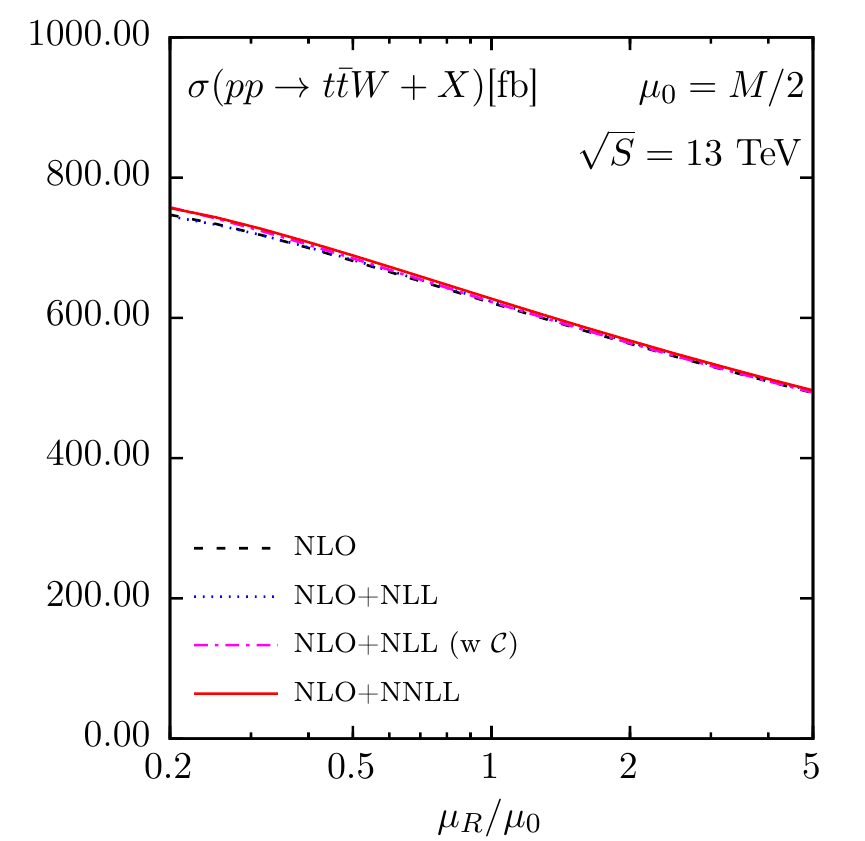}
\caption{Renormalisation scale dependence of the total cross section for the process $pp\to \ttW$ at the LHC with $\sqrt S=13$ TeV and $\muf=\mufo$ kept fixed. Results are shown for two central scale  values $\mu_0=\mufo=\muro=Q$ (left plot) and $\mu_0=\mufo=\muro=M/2$ (right plot).} 
\label{f:murdep_ttW}
\end{figure}

To further study the scale uncertainty of the total cross sections we show the dependence of the $\ttW$ and $\ttZ$ cross sections on the choice $\mu=\muf=\mur$ in figures~\ref{f:scaledependence_ttW}~and~\ref{f:scaledependence_ttZ}. For the associated production of the top quark pair with a $W$ boson, the sum of the $t \tb W^+$ and $t \tb W^-$ production is presented, since the two processes possess a very similar scale dependence. In figure~\ref{f:scaledependence_ttW}, a slight reduction in the scale dependence can be seen with the dominant contribution brought by  NLO+NLL~w~$\cal C$ result, indicating the importance of  contributions of hard origin. In addition, a mild increase of the dependence can be seen for the significantly small scales $\mu\lesssim0.3M/2$, which can potentially be attributed to the missing quark emission contribution. However, the scale at which this increase happens is not physically motivated and therefore of no relevance in practical studies. Separating the $\muf$ and $\mur$ dependence, i.e.\ varying the $\muf$ and $\mur$ while keeping  $\mur$ and $\muf$ fixed respectively,  leads to the conclusion that the $\ttW$ scale dependence is almost solely driven by the $\mur$ dependence, cf.\ figures~\ref{f:mufdep_ttW} and~\ref{f:murdep_ttW}.

\begin{figure}[h!]
\centering
\includegraphics[width=0.45\textwidth]{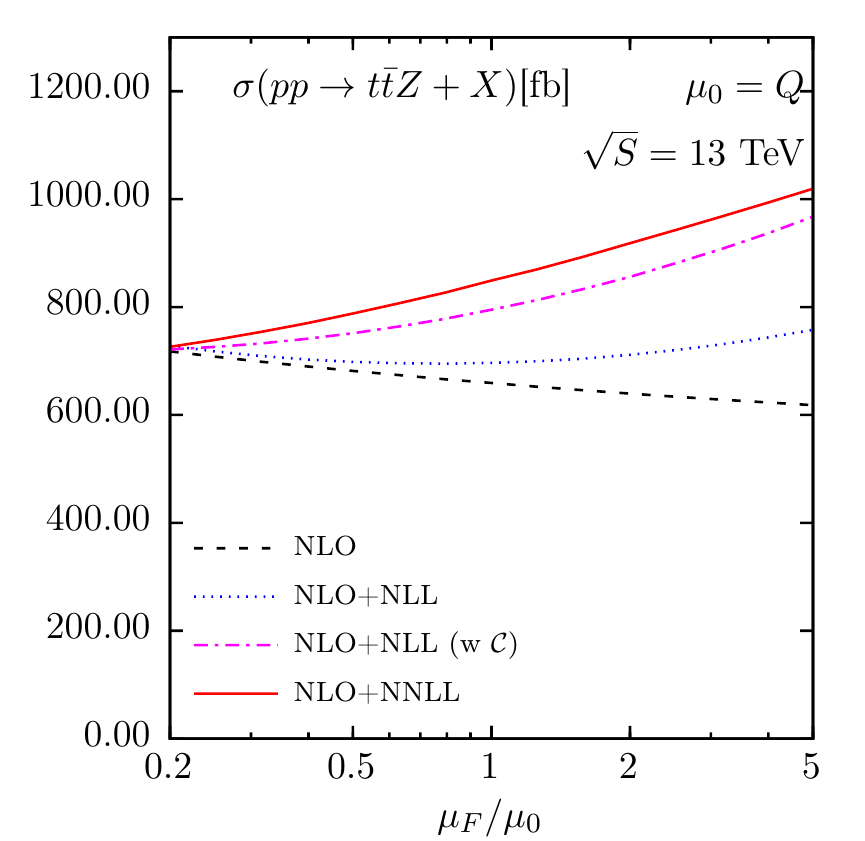}
\includegraphics[width=0.45\textwidth]{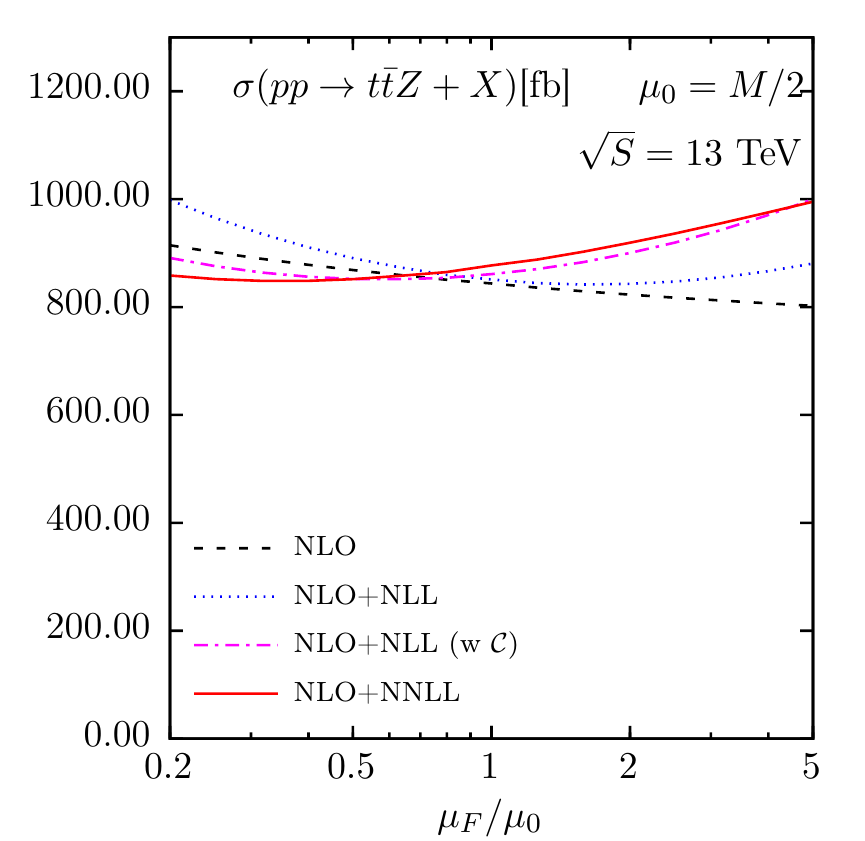}
\caption{Factorisation scale dependence of the total cross section for the process $pp\to \ttZ$ at the LHC with $\sqrt S=13$ TeV and $\mur=\mu_{R,0}$ kept fixed. Results are shown for  two central scale values $\mu_0=\mufo=\muro=Q$ (left plot) and $\mu_0=\mufo=\muro=M/2$ (right plot).} 
\label{f:murdep_ttZ}
\end{figure}

\begin{figure}[h!]
\centering
\includegraphics[width=0.45\textwidth]{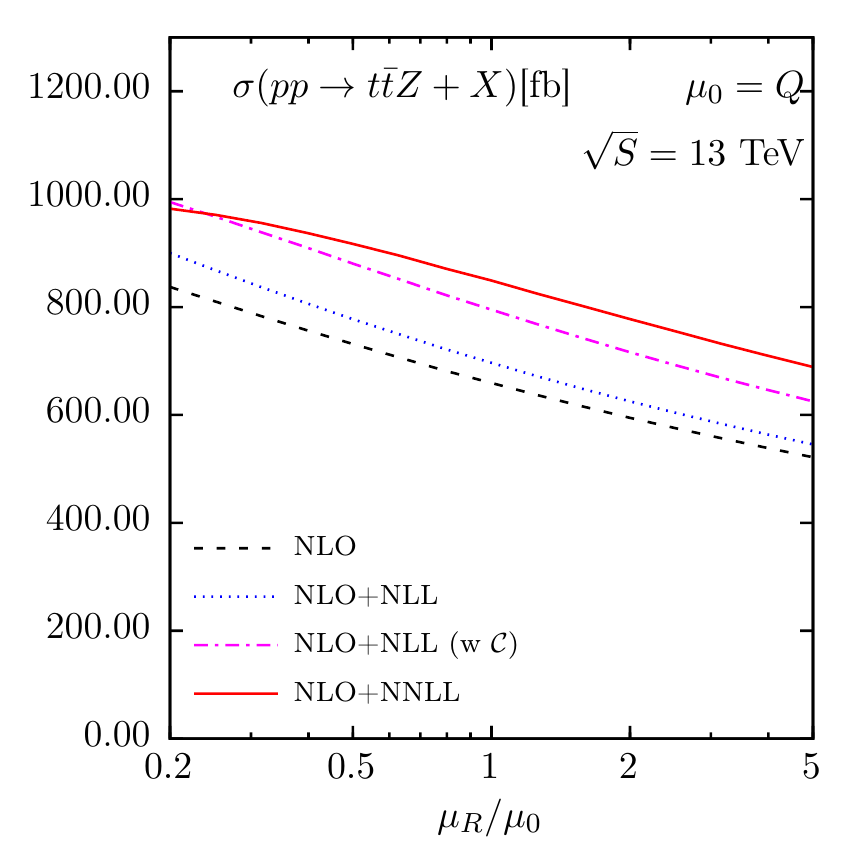}
\includegraphics[width=0.45\textwidth]{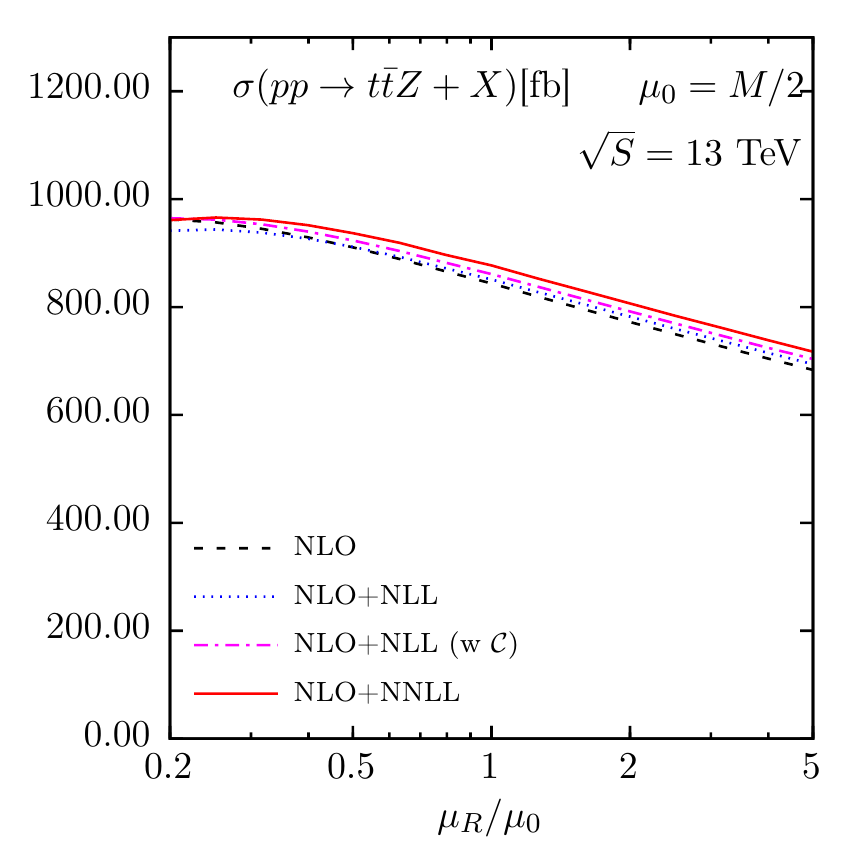}
\caption{Renormalisation scale dependence of the total cross section for the process $pp\to \ttZ$ at the LHC with $\sqrt S=13$ TeV and $\muf=\mufo$ kept fixed. Results are shown for two central scale values $\mu_0=\mufo=\muro=Q$ (left plot) and $\mu_0=\mufo=\muro=M/2$ (right plot).} 
\label{f:mufdep_ttZ}
\end{figure}

For the $\ttZ$ production process a more significant reduction in the dependence on $\mu=\muf=\mur$ can be seen in figure~\ref{f:scaledependence_ttZ}. Similarly to the $\ttW$ process, the dominant reduction in the uncertainty can also be attributed to the inclusion of constant contributions in $N$ from the hard and soft functions contained in the difference between the NLO+NLL and NLO+NLL~w~$\cal C$ results. However, a significant further reduction in the scale dependence originates from the resummation at NLL level and beyond. Additionally, the same increase of the dependence can be seen for the $\ttZ$ process at low scales, but again this effect concerns scales which choice is not physically motivated. The figures also illustrate that if we had attributed  the uncertainty of the cross section to the scale variation for $\muf=\mur$, the scale uncertainty would have been drastically reduced, even as low as to approximately 1\% for the $\mu=Q$ choice.  In contrast to the $\ttW$ process, the $\ttZ$  dependence on $\muf=\mur$ appears to be an effect of cancellations between dependencies on  $\muf$ and $\mur$. Taken separately they show an opposite behaviour, see figures~\ref{f:murdep_ttZ} and~\ref{f:mufdep_ttZ}. This behaviour is in fact very similar to the one observed for the process $pp \to \tth$, which also receives significant contributions from the $gg$ channel at LO.

\subsection{Invariant mass distributions}

\begin{figure}[h!]
\centering
\includegraphics[width=0.45\textwidth]{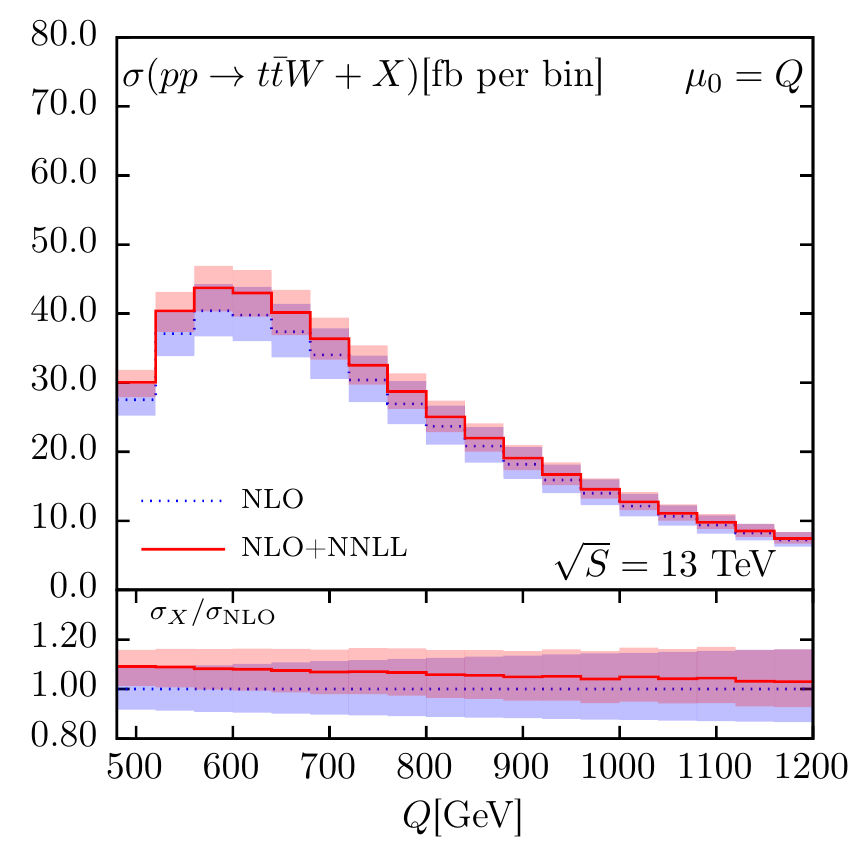}
\includegraphics[width=0.45\textwidth]{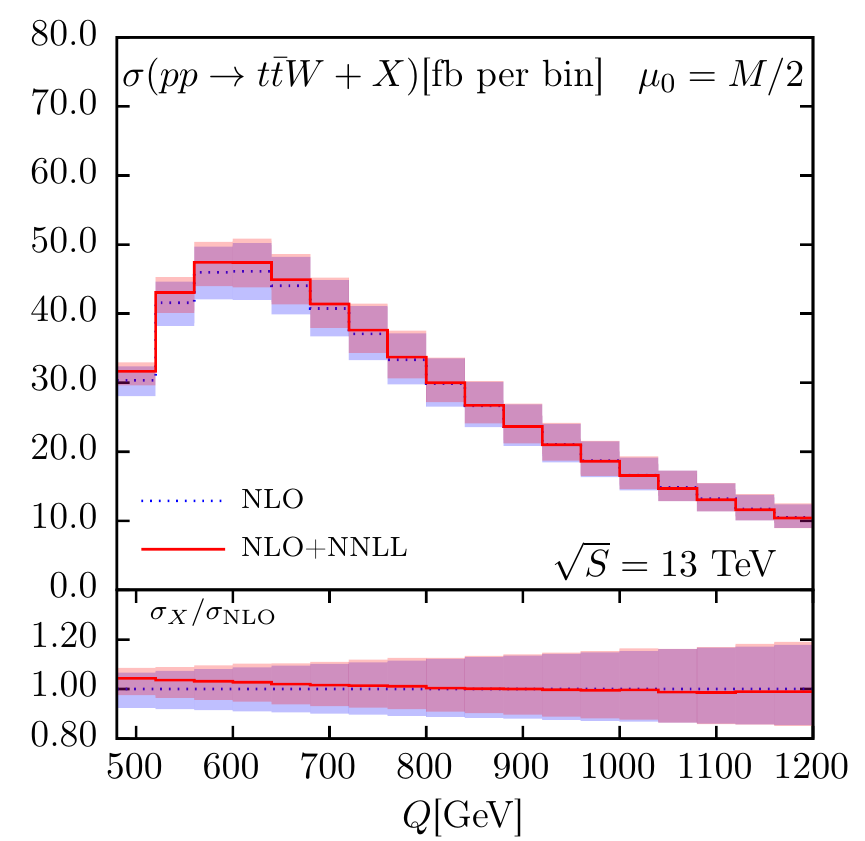}
\caption{Comparison of the NLO+NNLL and NLO invariant mass distributions for the process $pp \to \ttW$ at the LHC with $\sqrt S=13$ TeV. Results are shown for two central scale choices $\mu_0=Q$ (left plot) and $\mu_0=M/2$ (right plot). Lower panels show the ratio of the distributions w.r.t.\ the NLO predictions.} 
\label{f:Qdiff_ttW}
\end{figure}

\begin{figure}[h!]
\centering
\includegraphics[width=0.45\textwidth]{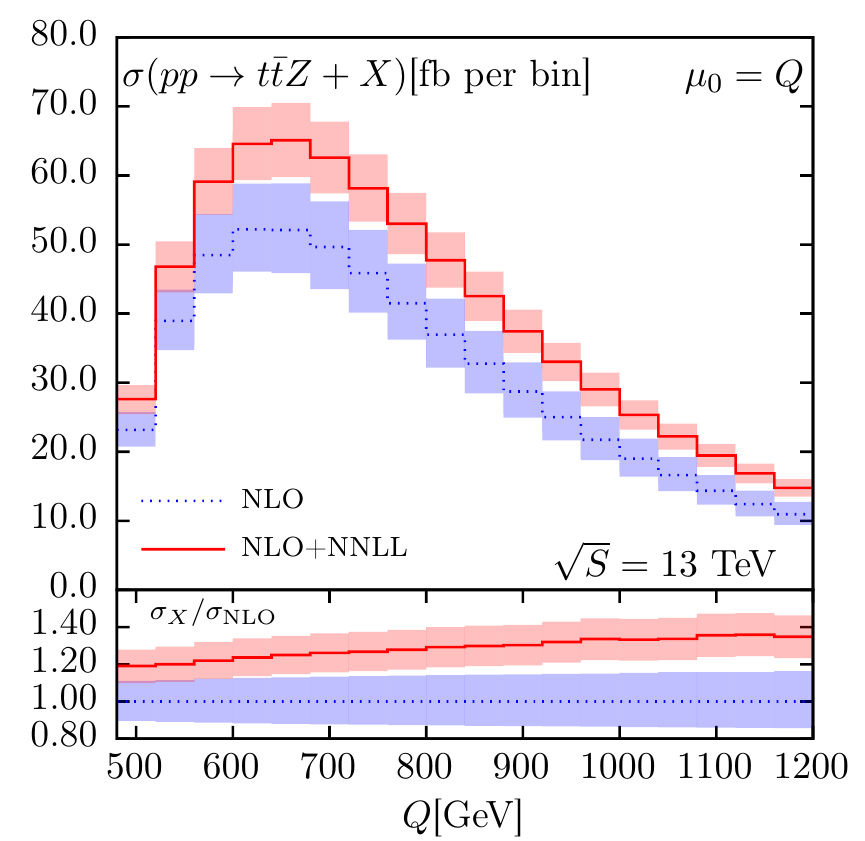}
\includegraphics[width=0.45\textwidth]{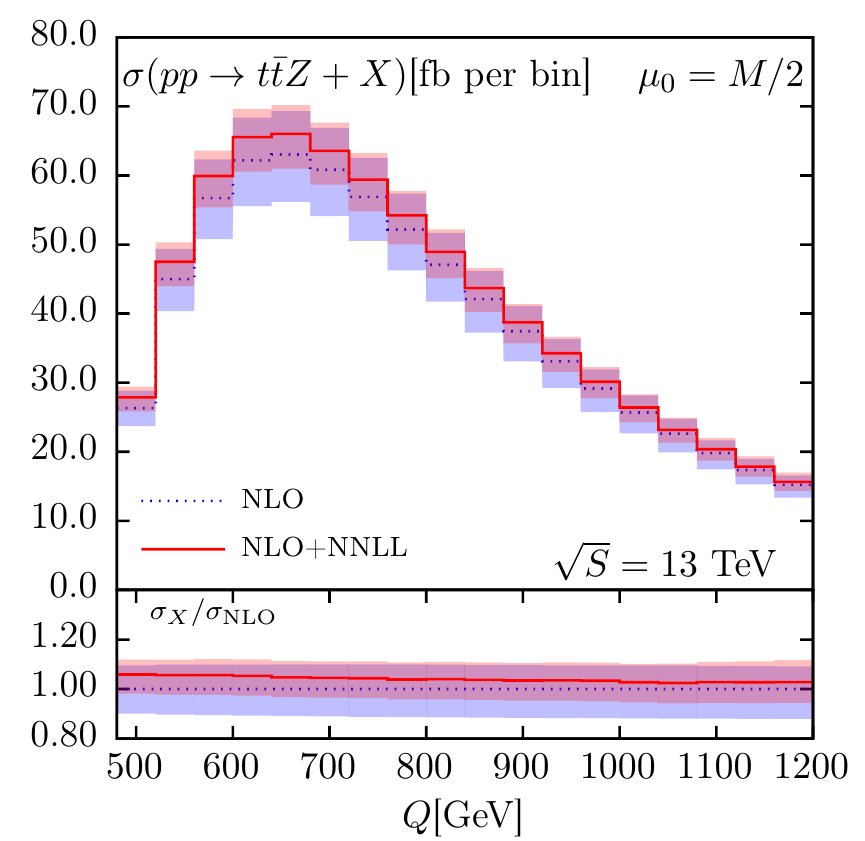}
\caption{Comparison of the NLO+NNLL and NLO invariant mass distributions for the process $pp \to \ttZ$ at the LHC with $\sqrt S=13$ TeV. Results are shown for two central scale choices $\mu_0=Q$ (left plot) and $\mu_0=M/2$ (right plot). Lower panels show the ratio of the distributions w.r.t.\ the NLO predictions.} 
\label{f:Qdiff_ttZ}
\end{figure}

\begin{figure}[h!]
\centering
\includegraphics[width=0.45\textwidth]{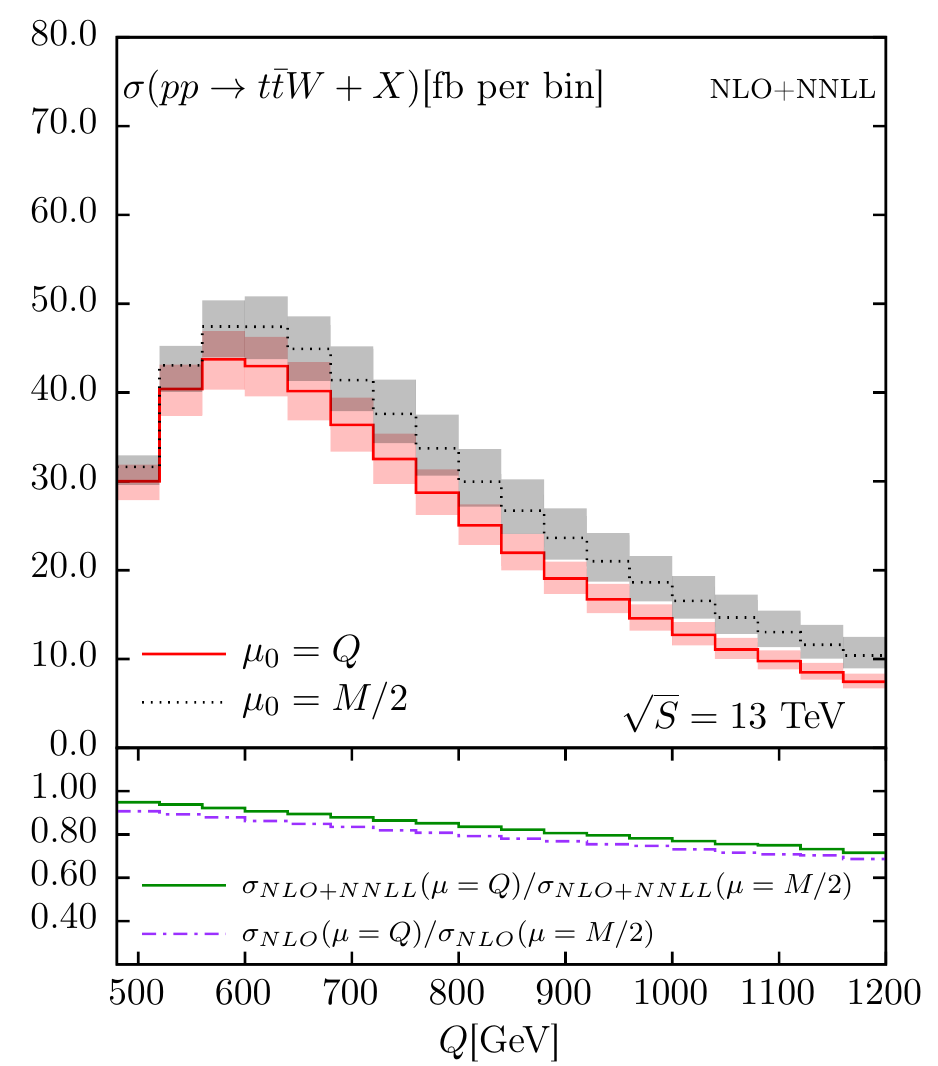}
\includegraphics[width=0.45\textwidth]{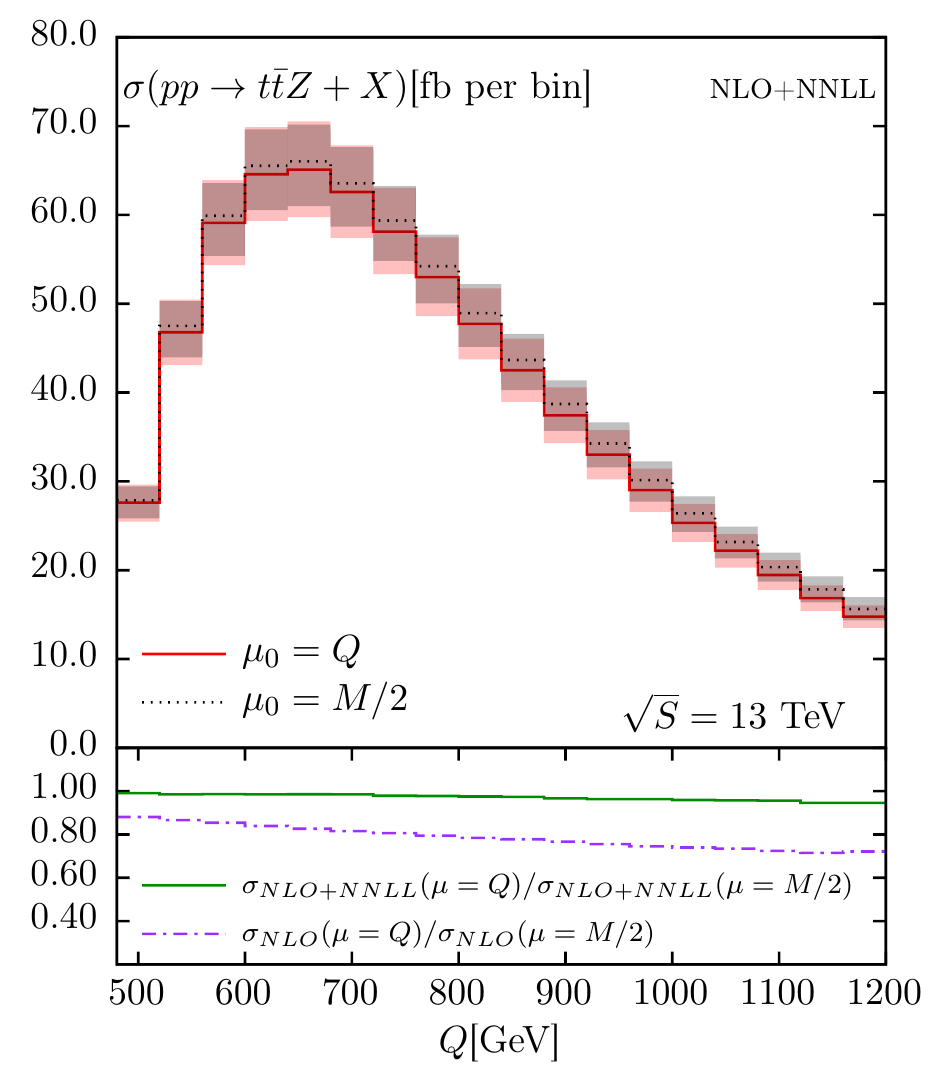}
\caption{Comparison of the NLO+NNLL invariant mass distributions for the process $pp \to \ttW$  (left plot) and  $pp \to \ttZ$ (right plot) at the LHC with $\sqrt S=13$ TeV. Results are shown for two central scale choices $\mu_0=Q$  and $\mu_0=M/2$. Lower panels show ratios of distributions calculated at either NLO or NLO+NNLL accuracy for these two scale choices.} 
\label{f:Qdiff_comp}
\end{figure}

Our total cross section predictions are obtained by integrating over invariant mass distributions $ {d\sigma}/{dQ^2}$. Note that these are the only distributions for which one has got a full control of the resummed contributions while performing threshold resummation in the invariant mass limit $\shat \to Q^2$.  The NLO+NNLL distributions in $Q$ for the two scale choices $\mu_0=Q$,  $\mu_0=M/2$ for the $\ttW$ and $\ttZ$ processes are presented in figures~\ref{f:Qdiff_ttW} and~\ref{f:Qdiff_ttZ}, respectively.  Apart from the scale choice, the size of the NNLL corrections depends now also on $Q$. In the $\ttW$ case, however, this dependence is moderate and the corrections do not exceed 10\%, cf.\ left plot in figure~\ref{f:Qdiff_ttW}. The corrections to the $\ttZ$ invariant mass distribution, on the other hand, show much stronger $Q$ dependence. Figure~\ref{f:Qdiff_ttZ} illustrates that the NNLL corrections can reach up to 40\% for the $\mu_0=Q$ scale choice which is a much higher value than the 29\% reported for the total cross section in table~\ref{t:totalxsec_13TeV}. Similarly as in the case of total cross sections, also for differential distributions inclusion of the NNLL corrections results in a much better agreement between theoretical predictions obtained with various scale choices, and in consequence leads to stabilization of the predictions, see figure~\ref{f:Qdiff_comp}.

\subsection{Comparison with other NLO+NNLL predictions in the literature}

The NLO+NNLL predictions for the associated $\ttW$ and $\ttZ$ production calculated in the SCET framework are available~\cite{Li:2014ula, Broggio:2016zgg,Broggio:2017kzi}. By comparing our results obtained using the direct QCD approach, we can not only deliver an independent check of  the previously published result but also gain insights on the size of the subleading i.e.  below formal accuracy effects which are  treated differently by the two methods.  In order to perform the corresponding comparisons we used the same values of parameters and the same pdf sets as in the above mentioned papers. It has to be noted though, that the scale choices made to obtain results reported in
this paper and~\cite{Li:2014ula, Broggio:2016zgg, Broggio:2017kzi}  are not equivalent. While our resummed expressions depend on
$\muf$ and $\mur$, the formulas obtained in the SCET formalism contain dependence on the hard and soft scales
$\mu_h$ and $\mu_s$, as well as $\muf$. In particular, Ref.~\cite{Li:2014ula}  uses the choice $\muf = \mu_h=Q$ and a minimizing procedure to set $\mu_s$. Nevertheless, we find a very good agreement with our results calculated using the scale choice $\muf =\mur =Q$. Specifically, the authors ~\cite{Li:2014ula} obtain $\sigma_{\rm NLO+NNLL}=332.99^{+5\%}_{-4\%}$ fb for the $\ttW^+$ production and $\sigma_{\rm NLO+NNLL}=169.86^{+5\%}_{-4\%}$ fb for the $\ttW^-$ production at $\sqrt S=13$ TeV, while we have $\sigma_{\rm NLO+NNLL}=331^{+8.9\%}_{-8.6\%}$ fb and $\sigma_{\rm NLO+NNLL}=170^{+8.8\%}_{-8.6\%}$ fb, correspondingly.

Ref.~\cite{Broggio:2016zgg} also reports the NLO+NNLL predictions for the $\ttWp$ and $\ttWm$ processes at the LHC. Contrary to~\cite{Li:2014ula}, the  $\mu_s$ scale in~\cite{Broggio:2016zgg} is chosen in such a way as to mimic the scale of soft radiation in the Mellin-space framework, i.e. $\mu_ s = Q/N$. As pointed out in~\cite{Kulesza:2017ukk}, only one choice $\muf = \mu_h=Q$ and $\mu_s=Q/N$ in the scale setting procedure of~\cite{Broggio:2016zgg} directly corresponds to  setting $\mu = \muf = \mur=Q$ in our results. With this choice, and using the same pdf and input parameter setup as in~\cite{Broggio:2016zgg}, we obtain $\sigma_{\rm NLO+NNLL}=328.6^{+29.2}_{-28.1}$~fb  for the $\ttWp$ and $\sigma_{\rm NLO+NNLL}=171.2^{+15.0}_{-14.6}$~fb for the $\ttWm$ process at $\sqrt S$=13 TeV, to be compared with $\sigma_{\rm NLO+NNLL}=333^{+14.9}_{-12.4}$ fb  and  $\sigma_{\rm NLO+NNLL}=173.1^{+7.7}_{-6.6}$ fb  reported in~\cite{Broggio:2016zgg}. The differences between the central values obtained by these two different calculations amount thus to 1\%. Note that, similarly to~\cite{Li:2014ula}, scale errors cannot be directly compared due to different methods used for calculating them. In particular, as explained above, our estimates of the scale error are calculated using the seven point method.

\begin{figure}[h!]
\centering
\includegraphics[width=0.45\textwidth]{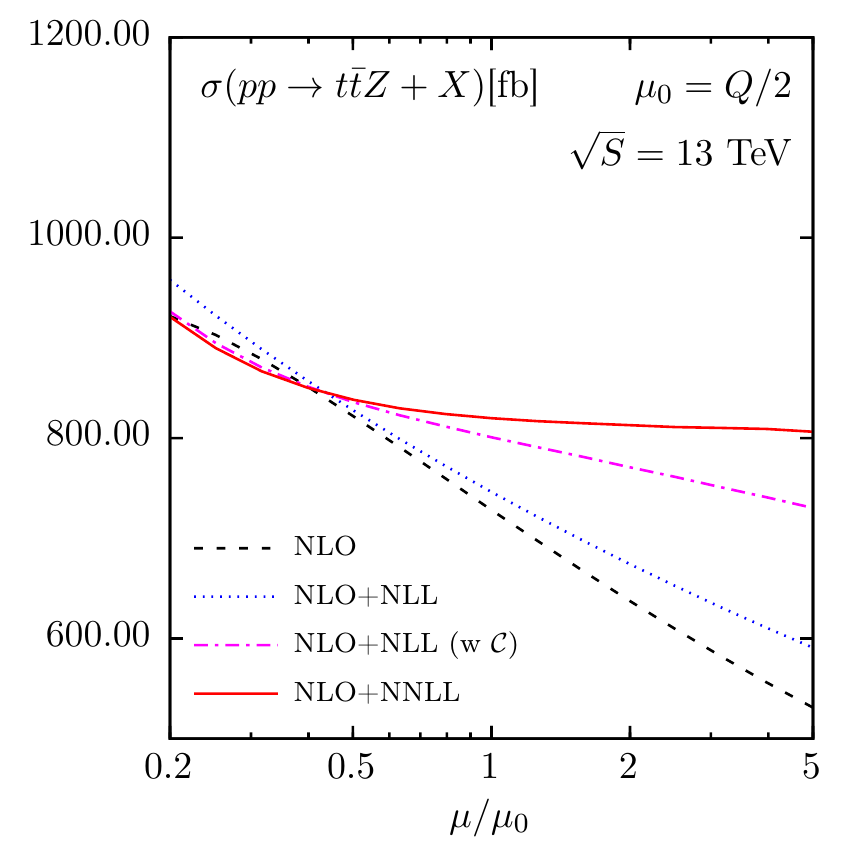}
\caption{Scale dependence of the total cross section for the process $pp\to \ttZ$ at the LHC with $\sqrt S=13$ TeV, calculated with MMHT2014 pdf set and the input parameter values for~\cite{Broggio:2017kzi}. Results are shown for the choice $\mu=\muf=\mur$ and the central scale value $\mu_0=Q/2$.} 
\label{f:mudep_ttZ_SCET}
\end{figure}

The NLO+NNLL predictions for the $\ttZ$ production process reported in table 2 of~\cite{Broggio:2017kzi} use yet another scale setup with $\muf = Q/2$, $\mu_h=Q$ and $\mu_ s = Q/N$. Therefore these predictions cannot be directly compared with ours. Nevertheless, in figure~\ref{f:mudep_ttZ_SCET} we present our NLO+NNL, NLO+NLL (w $\cal{C}$), NLO+NNLL results as well as the NLO cross section for the $\ttZ$ production using the same pdf and input parameter setup as in~\cite{Broggio:2017kzi} and choosing $\mu_0=\mufo=\muro =Q/2$. Notably, within the range of the scales considered, we do not see the rising behaviour of the NLO+NNLL cross sections with the growing scale as in figure~1 of ~\cite{Broggio:2017kzi}, instead a relatively flat one, especially at $\mu \leq Q/4$. Also, as discussed in Section~\ref{s:totalxs}, compared to NLO+NLL predictions our NLO+NNLL results show decreased dependence on the scale variation. Regarding numerical values for $\sqrt S$=13 TeV, we obtain $\sigma_{\rm NLO+NNLL}=819.7^{+58.6}_{-65.0}$~fb at the scale $\mu=\muf=\mur=Q/2$  whereas~\cite{Broggio:2017kzi} reports  $\sigma_{\rm NLO+NNLL}=777.8^{+61.3}_{-65.2}$~fb. For comparison, the difference between the NLO+NNLL total cross section for the $\tth$ process obtained in the SCET formalism with $\muf = \mu_h=Q, \mu_s=Q/N$ and the NLO+NNLL total cross section obtained using the direct QCD method with $\muf = \mur=Q$~\cite{Kulesza:2017ukk}  amounts to 2.5\% at $\sqrt S$=13 TeV when the same pdf and input parameter setup is used. Given this, and the values we can read off from figure~1 of~\cite{Broggio:2017kzi}, we expect the difference between the two cross sections quoted here to be significantly impacted by the scale setting procedure.

\subsection{Comparison with the $\ttV$  total cross section measurements at the LHC}

The currently most precise measurements of $t\bar t W^{\pm}$ and $t\bar t Z$ cross sections in $pp$ collisions at $\sqrt{S} = 13$~TeV were recently published by CMS \cite{Sirunyan:2017uzs} and ATLAS collaborations \cite{ATLAS:2018ekx} . The data samples correspond to an integrated luminosity of  35.9~fb$^{-1}$ and 36.1~fb$^{-1}$, respectively. While ATLAS measures $t\bar t W^{\pm}$ and $t\bar t Z$ production cross sections simultaneously, CMS provides numerical values for individual measurements and a figure with the results of a simultaneous fit. In table~\ref{t:exp_vs_theory_13TeV} we compare the results of these measurements with the  central theoretical predictions of this paper at the NLO+NNLL accuracy (\ref{eq:totalxsec_ttZ_combined}), (\ref{eq:totalxsec_ttWp_combined}) and (\ref{eq:totalxsec_ttWm_combined}), to which electroweak corrections reported in~\cite{deFlorian:2016spz} are added. For each process, the EW corrections are estimated from the values of the relative EW corrections listed in table 40 of~\cite{deFlorian:2016spz} ($-0.2$\% for $\ttZ$, $-3.5$\% for $\ttWp$, $-2.6$\% for $\ttWm$, $-3.2$\% for $\ttW$),  and the corresponding NLO QCD cross sections calculated using the envelope method and the NLO values listed in table 1 in this paper.  The QCD uncertainties are applied also to the EW correction effects. In this way we provide theoretical predictions which include state-of-the-art knowledge of the QCD at the NLO+NNLL accuracy and the EW effects up to the NLO accuracy.

\begin{table}[h!]
	\begin{center}
\renewcommand{\arraystretch}{1.5}
		\begin{tabular}{| l | l | l |}
			\hline
process &
experiment &
NLO + NNLL + EW \tabularnewline
& $\sigma$ $\pm$ stat.\ err.\ $\pm$ syst.\ err. [pb]
& $\sigma$ $\pm$ scale err.\ $\pm$ pdf+$\als$ err. [pb]
\tabularnewline
			\hline
$t \bar t W^{+}$   &
$0.58 \pm 0.09 {}^{+0.09}  _{-0.08}$ \ \ (CMS)&
$0.36 ^{+0.09} _{-0.06} \pm {0.01}$
\tabularnewline
$t \bar t W^{-}$   &
$0.19 \pm 0.07 \pm 0.06 $\ \  (CMS)
&
$0.19 ^{+0.05} _{-0.03} \pm {0.01}$ 
\tabularnewline
$t \bar t W$   &
$0.77^ {+0.12}_{-0.11}\;{} ^{+ 0.13}_{-0.12} $\ \  (CMS)
&
$0.55 ^{+0.14} _{-0.09} \pm {0.02}$
\tabularnewline
$t \bar t W$   &
$0.87 \pm 0.13 \pm 0.14 $\ \  (ATLAS)
&
$0.55 ^{+0.14} _{-0.09} \pm {0.02}$ 
\tabularnewline
\hline
$t \bar t Z$    &
$0.99^{+0.09} _{-0.08}\; {}^{+ 0.12} _{-0.10}$\ \ (CMS) 
&
$0.86 ^{+0.07 } _{-0.08} \pm {0.03}$ 
\tabularnewline
$t \bar t Z$   &
$0.95 \pm 0.08 \pm 0.10 $ \ \ (ATLAS)& 
$0.86 ^{+0.07 } _{-0.08} \pm {0.03}$ 
\tabularnewline
			\hline
		\end{tabular}
	\end{center}
\caption{Results of experimental measurements
by  CMS~\cite{Sirunyan:2017uzs} and ATLAS~\cite{ATLAS:2018ekx} collaborations
of total cross sections $\sigma$ for $pp \to t \bar t W^{+}/W^{-}/Z$ at $\sqrt S=13$ TeV compared to theory predictions at NLO+NNLL accuracy given in eqs.~(\ref{eq:totalxsec_ttZ_combined}), ~(\ref{eq:totalxsec_ttWp_combined}) and ~(\ref{eq:totalxsec_ttWm_combined}) with added the electroweak corrections as reported in~\cite{deFlorian:2016spz}.  The scale and the pdf+$\als$ errors correspond to QCD cross sections.}
\label{t:exp_vs_theory_13TeV}
\end{table}

The comparison shows good agreement between theory and data within errors. The largest relative difference between theory and experiment is found for  the CMS measurement of the $t \bar t W^+$ production. Even with the most conservative estimate of scale error provided by the envelope method, the overall theory error are smaller than the current experimental errors. It is interesting to note a general trend for the theoretical results to be lower or comparable to the measured experimental values. The same conclusions hold if instead the NLO+NNLL predictions with the conservative scale error estimates provided by the envelope method, NLO+NNLL predictions for the scale choice  $\mu=M/2$ are considered.

\begin{figure}[t!]
\centering
\includegraphics[width=0.45\textwidth]{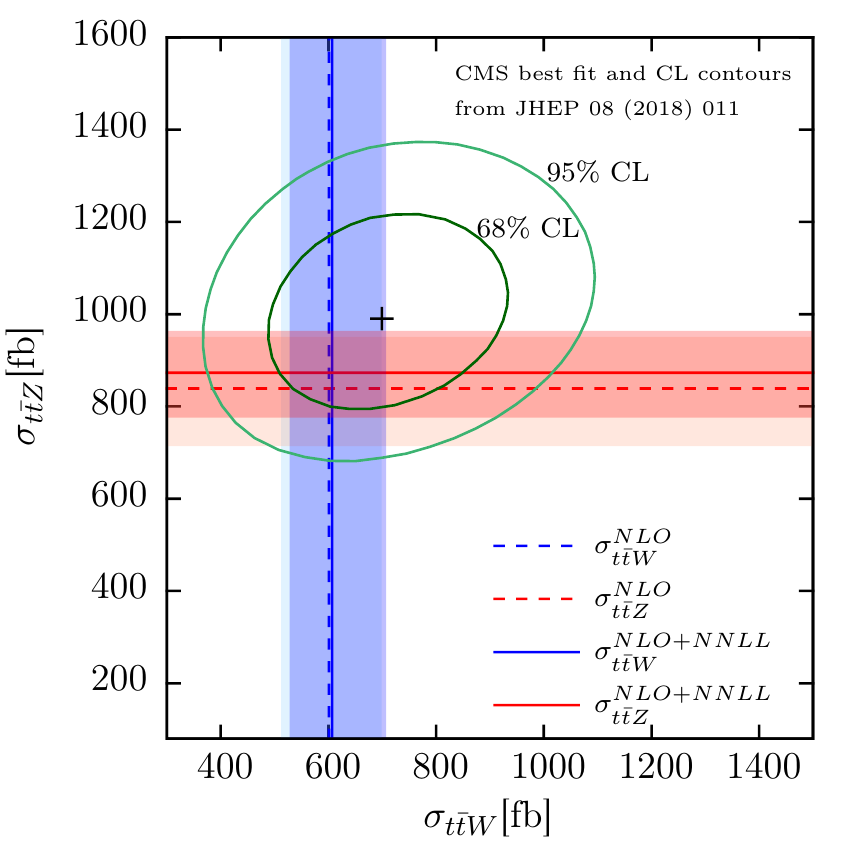}
\includegraphics[width=0.45\textwidth]{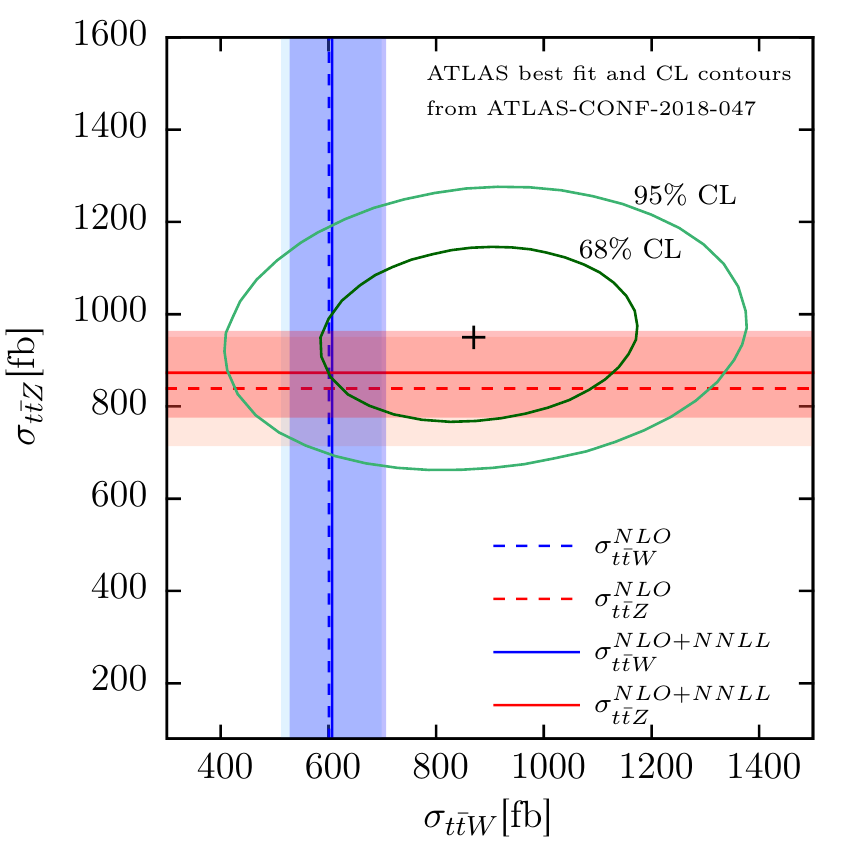}
\caption{NLO+NNLL and NLO predictions for the total $\ttZ$ and $\ttW$ cross sections at the central scale  $\mufo=\muro=M/2$  and with added electroweak corrections reported in~\cite{deFlorian:2016spz}, compared to the CMS~\cite{Sirunyan:2017uzs}
 (left plot) and ATLAS~\cite{ATLAS:2018ekx} (right plot) measurements. } 
\label{f:totalxsec_exp}
\end{figure}
\begin{figure}[h!]
\centering
\includegraphics[width=0.45\textwidth]{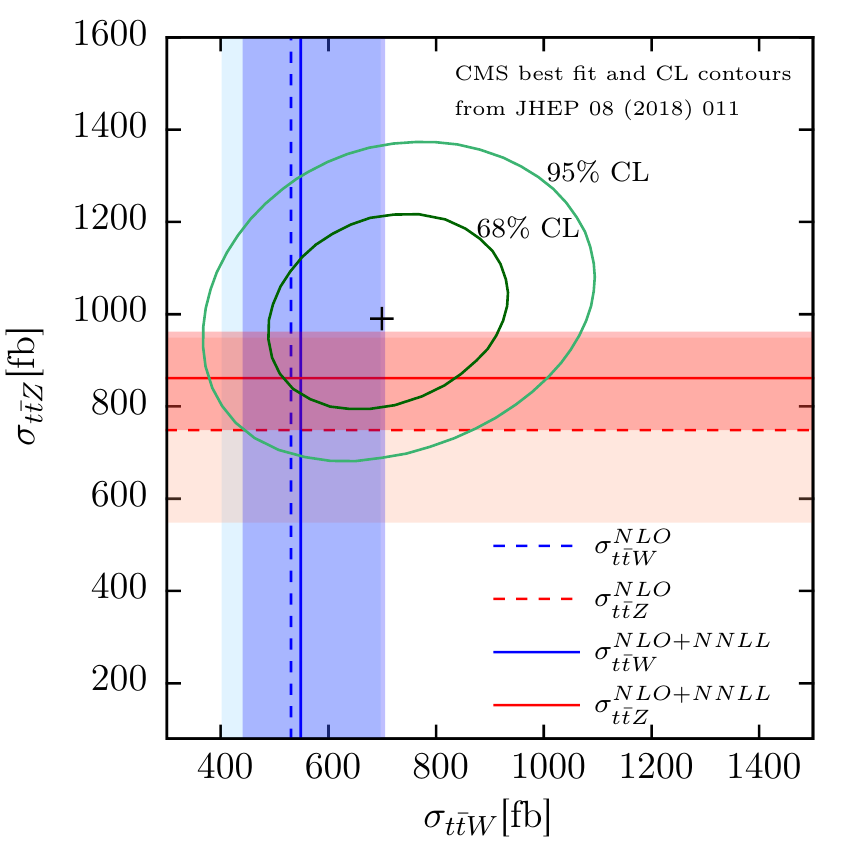}
\includegraphics[width=0.45\textwidth]{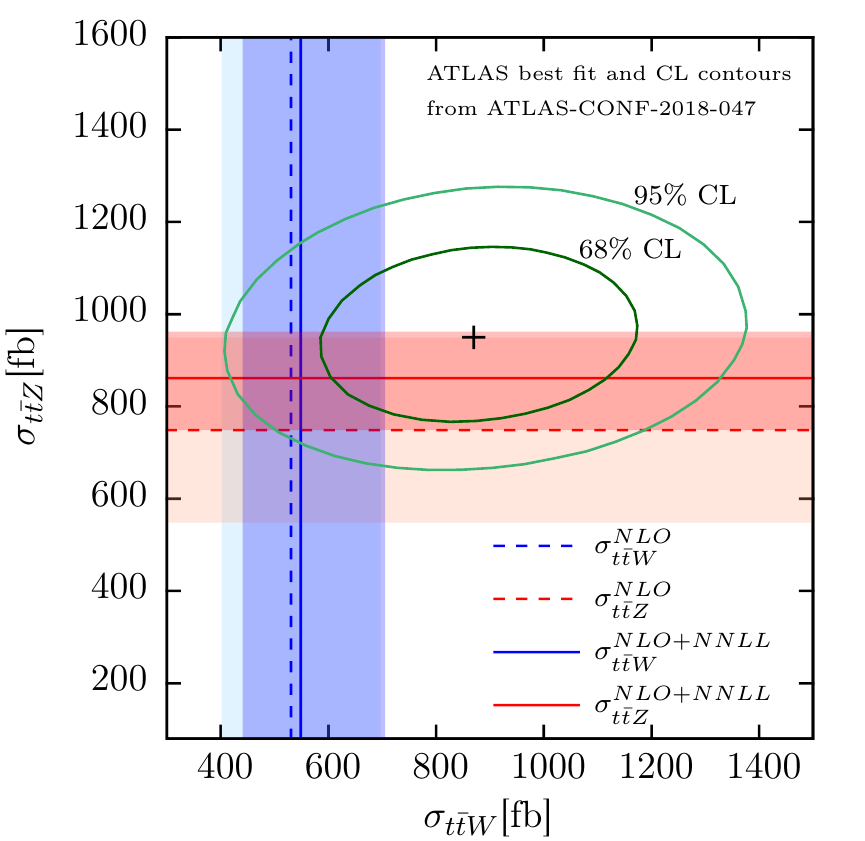}
\caption{NLO+NNLL and NLO predictions for the total $\ttZ$ and $\ttW$ cross sections, using the envelope method as described in the text, and with added  electroweak corrections reported in~\cite{deFlorian:2016spz},  compared to the CMS~\cite{Sirunyan:2017uzs}
 (left plot) and ATLAS~\cite{ATLAS:2018ekx} (right plot) measurements. } 
\label{f:totalxsec_exp_avg}
\end{figure}

In addition, in figures\ \ref{f:totalxsec_exp} and\ \ref{f:totalxsec_exp_avg} we show the NLO predictions and our NLO+NNLL predictions for the $t\bar t Z$ and $t\bar t W$ processes, to which we add the electroweak corrections computed in~\cite{deFlorian:2016spz}. The NLO+NNLL results are marked by full lines for the central values and darker shaded bands for the errors, while for NLO dashed lines and light shaded bands are used correspondingly.  
In the figures we also plot the values of the corresponding cross sections measured using a combined fit by the CMS~\cite{Sirunyan:2017uzs} and ATLAS~\cite{ATLAS:2018ekx} collaborations, together with their confidence level (CL) contours. The theory errors are estimated by adding the the scale errors and pdf+$\als$ errors of the QCD cross sections. In figure\ \ref{f:totalxsec_exp} the theory values for the central scale $\mu_0 = M/2$ are shown. The NLO QCD+EW results with this scale choice have been reported in the Yellow Report~\cite{deFlorian:2016spz}, and are taken as the benchmark theory predictions by the experiments. 
Since, however, the choice of a single fixed scale at the NLO accuracy may lead to underestimation of the theoretical uncertainty, in figure\ \ref{f:totalxsec_exp_avg} we display also the NLO+NNLL and NLO predictions that include the whole span of the central scales from $M/2$ to $Q$. The NLO+NNLL results~(\ref{eq:totalxsec_ttZ_combined}),~(\ref{eq:totalxsec_ttWp_combined}) and~(\ref{eq:totalxsec_ttWm_combined}) with added EW corrections combine predictions for various central scale choices, thus  yielding more conservative error estimates. 
The two-dimensional analyses visualised in figures~\ref{f:totalxsec_exp} and~\ref{f:totalxsec_exp_avg} confirm good agreement between the theory predictions and the measurements by the ATLAS and the CMS collaborations. Since the NLO+NNLL total cross sections are higher than the NLO ones, the NNLL calculations result in bringing the central values of the theoretical predictions closer to the experimentally measured cross sections. In the case of analysis with more conservative error estimates (figure~\ref{f:totalxsec_exp_avg}), this distance gets reduced by as much as a factor of two for the $t\bar t Z$ production. The theoretical accuracy for $t\bar t Z$ in this conservative approach is equally well improved due to inclusion of the soft gluon resummation effects in the NNLL approximation, also by around a factor of two w.r.t.\ the NLO result.

\section{Summary}

In this paper soft gluon corrections were calculated to $\ttV$ production in association with a heavy electroweak boson $V$, $V=W^{\pm}$ or $Z$ in $pp$ collisions. The calculations were performed in the three particle invariant mass kinematics through NNLL accuracy, and the results were matched to existing NLO results. Resummation was achieved using the direct QCD approach in the Mellin space. We computed invariant mass distributions and the total cross sections, obtained by integration of these distributions. In particular, we calculated NLO+NNLL total cross sections  for the LHC collisions at $\sqrt{S} =13$~TeV and $\sqrt{S} = 14$~TeV.  The scale uncertainty of these predictions were estimated by independent variation of the factorisation and renormalisation scales around a central scale $\mu_0$, using the seven point method. Three different choices of the central scale $\mu_0$ were assumed:  $\mu_0 = M/2$, $\mu_0 = Q/2$  and $\mu_0 = Q$, where $M = 2 m_t + m_V$ is the absolute threshold energy, and $Q$ is the invariant mass of the  $t \bar t V$ system. 

The effect of soft gluon corrections was found to be more important for the $t \bar t Z$ than for the $t \bar t W^{\pm}$ production. This was expected, as the LO amplitudes for  $t \bar t W^{\pm}$ production are driven by quark scattering, while for  $t \bar t Z$ the two gluon production channel is dominant, and stronger gluon radiation occurs due to higher colour charges.  For the $t \bar t Z$ production we observed a substantial improvement of the theoretical accuracy due to inclusion of the soft gluon corrections. First of all, the results are much more stable w.r.t.\ the central scale choice: at NLO+NNLL the total cross section increases by only 3\% when $\mu_0$ decreases from $Q$ to $M/2$, while the corresponding increase is 28\% at NLO. Moreover, for the fixed central scale, the dominant theory errors from the scale choice uncertainty decrease by 29\% -- 38\% by going from NLO to NLO+NNLL.  A conservative estimate of the theoretical accuracy obtained as an envelope over results for various scale choices and  their errors is improved by up to a factor of two by performing the NNLL soft gluon resummation. As in the case of the Higgs boson production with association of $t \bar t$ quarks, our results are compatible with NLO prediction for the central scale choice $\mu_0 = \mufo=\muro= M/2$  justifying that common choice at least for the $\ttZ$ process.

The obtained results were compared to the existing predictions for the $t \bar t W^{\pm}$ and $t\bar t Z$ cross sections  at NLO+NNLL that were calculated in the SCET framework. In order to perform a meaningful comparison, we computed the cross sections employing the same sets of parton distribution functions and the input parameters as in those papers. For equivalent scale choice setups,  our NLO+NNLL predictions and the cross sections calculated using the SCET framework agree well.

Finally, the theoretical estimates of $t\bar t V$ total cross sections were compared to the latest ATLAS and CMS measurements  at $\sqrt{S} =13$~TeV. A good agreement was found between theory and data. In a two dimensional analysis of $t\bar t W$ and $t \bar t Z$ cross sections, the combined experimental data differ by about one standard deviation from the results of this paper. In comparison with the NLO predictions, the NLO+NNLL calculations result in theoretical predictions with central values closer to the measured experimental cross sections. The errors of the NLO+NNLL predictions are in general smaller than the current experimental errors.

\section*{Acknowledgments}
We thank V.~Hirschi and M.~Zaro for the correspondence regarding MadGraph5\_aMC@NLO $\ttV$ results published in the HXSWG Yellow Report 4.
This work has been supported in part by the DFG Grant KU 3103/1 and by Polish National Science Center (NCN) grant No.\ 2017/27/B/ST2/02755.
TS would like to thank the Ministry of Science and Higher Education of Poland for support in the form of the Mobility Plus grant as well as Brookhaven National Laboratory for hospitality and support. 
The work leading to this publication was also supported by the German Academic Exchange Service (DAAD) with funds from the German Federal Ministry of Education and Research (BMBF) and the People Programme (Marie Curie Actions) of the European Union Seventh Framework Programme (FP7/2007-2013) under REA grant agreement n.\ 605728 (P.R.I.M.E.\ Postdoctoral Researchers International Mobility Experience).
AK would like to acknowledge the Mainz Institute for Theoretical Physics for hospitality and its support during a part of work on this project.
 
\bibliographystyle{JHEP}
\providecommand{\href}[2]{#2}\begingroup\raggedright\endgroup

\end{document}